%% file: hcr.tex
\documentclass[iop]{emulateapj}

\usepackage{graphicx}
\usepackage{lineno}                   
\usepackage{amsmath}                  
\usepackage{dcolumn}                  
\usepackage{bm}                       
\usepackage{xspace}                   
\usepackage{hyperref}                 

\shorttitle{Observation of Cosmic-Ray Anisotropy with HAWC}
\shortauthors{HAWC Collaboration}

\begin{document}
\title{Observation of Small-scale Anisotropy
in the Arrival Direction Distribution\\
of TeV Cosmic Rays with HAWC }

\input authorsapj

\begin{abstract}
The High-Altitude Water Cherenkov (HAWC) Observatory is sensitive to
gamma rays and charged cosmic rays at TeV energies.  The detector is still
under construction, but data acquisition with the partially deployed detector
started in 2013.  An analysis of the cosmic-ray arrival direction distribution
based on $4.9\times 10^{10}$ events recorded between June 2013 and February 
2014 shows anisotropy at the $10^{-4}$ level on angular scales of about 
$10^\circ$.  The HAWC cosmic-ray sky map exhibits three regions of significantly
enhanced cosmic-ray flux; two of these regions were first reported by 
the Milagro experiment.  A third region coincides with an excess recently 
reported by the ARGO-YBJ experiment.  An angular power spectrum analysis of the 
sky shows that all terms up to $\ell=15$ contribute significantly to the excesses.  
\end{abstract}


\section{Introduction}
\label{sec:Introduction}

The High-Altitude Water Cherenkov (HAWC) Observatory is designed to study the
sky in gamma rays and cosmic rays between 50\,GeV and 100\,TeV.  The detector
is currently under construction 4100\,m above sea level at the saddle point
between Volc\'{a}n Sierra Negra and Pico de Orizaba near Puebla, Mexico, at
$19^\circ$N latitude.  HAWC is a water-Cherenkov extensive air-shower array
with a wide field of view and nearly 100\,\% duty cycle.  With its daily sky
coverage of $8.4$\,sr, HAWC will record both steady and transient gamma-ray
sources and provide an unbiased survey of the sky between $-26^\circ$ and
$64^\circ$ in declination.  While the main targets of HAWC are gamma-ray
sources, the detector is also sensitive to cosmic rays.  The large number of
cosmic rays detected with HAWC forms an undesirable background in the search for
gamma-ray sources, but it also permits precise measurements of small
deviations from isotropy in the cosmic-ray flux at TeV energies.

Anisotropy in the arrival direction distribution of TeV cosmic rays has been
observed with detectors in the Northern and Southern Hemispheres.  In the
northern sky it has been measured with the Tibet AS$\gamma$~\citep{Amenomori:2005dy}, 
Super-Kamiokande~\citep{Guillian:2007}, Milagro~\citep{Abdo:2008kr,Abdo:2008aw}, 
EAS-TOP~\citep{Aglietta:2009mu}, MINOS~\citep{deJong:2012hk}, and 
ARGO-YBJ~\citep{DiSciascio:2013cia, ARGO-YBJ:2013gya} experiments.  
In the Southern Hemisphere, the only measurements come from the IceCube
detector~\citep{Abbasi:2010mf,Abbasi:2011ai,Abbasi:2011zka} and its surface air
shower array, IceTop~\citep{Aartsen:2012ma}.  Observations in the northern and
southern sky show qualitatively similar results.  In both hemispheres, the
anisotropy has components on large angular scales ($>60^\circ$) and on smaller
scales ($<60^\circ$).  The large-scale anisotropy is dominated by an approximately 
dipole structure with amplitude of order $10^{-3}$ in relative intensity which 
persists up to at least 2 PeV~\citep{Aartsen:2012ma}, although the dipole phase is 
observed to change above 100\,TeV~\citep{Abbasi:2010mf,Aglietta:2009mu}.
The small-scale structure ranges in relative intensity from several $10^{-4}$ 
to $10^{-3}$.

The anisotropy in the cosmic-ray flux at these energies is not well-understood.
The Larmor radius of a TeV proton in a $\mu$G magnetic field is approximately
0.001\,pc, orders of magnitude less than the distance to potential astrophysical 
accelerators, so cosmic rays from these sources should not point back to their 
origin.  It has long been suggested that weak dipole or dipole-like features 
should be a consequence of the diffusion of cosmic rays from nearby sources in the
Galaxy~\citep{Erlykin:2006ri,Blasi:2011fm,Pohl:2012xs,Sveshnikova:2013ui}. It
is also possible, though not yet demonstrated, that the magnetic fields of the 
heliosphere have an influence on the anisotropy~\citep{Desiati:2011xg,Schwadron:2014}. 
The small-scale structure, on the other hand, could be the product of turbulence in
the Galactic magnetic field~\citep{Giacinti:2011mz,Ahlers:2013ima} or an
additional heliospheric effect~\citep{Drury:2013uka}.  Several authors have
also suggested that the small-scale structure is produced in the decay of quark
matter present in pulsars~\citep{Perez-Garcia:2013lza} or in the
self-annihilation of dark matter~\citep{Harding:2013qra}.

Data acquisition with HAWC started in June 2013, and since then the instrument
has accumulated a data set that is already sufficiently large to study
cosmic-ray anisotropy at the $10^{-4}$ level in relative intensity.  HAWC data
cover a part of the sky that has been extensively studied by the Milagro
experiment, which operated near Los Alamos, New Mexico, between 2000 and 2008,
and the Tibet AS$\gamma$ and ARGO-YBJ experiments.  HAWC also slightly extends
the sky coverage of the previous measurements to declinations as low as
$-26^\circ$, observing heretofore uncharted declinations and narrowing the gap
in sky coverage between the Northern Hemisphere measurements and those
performed with IceCube at the South Pole.  The median energy of cosmic rays
observed by HAWC in the configuration used in this analysis is $\sim2$\,TeV, 
comparable to Milagro (1\,TeV) and ARGO-YBJ (1.8\,TeV).  The actual energy 
distribution of the events is likely to be more similar in HAWC and ARGO-YBJ, 
as both detectors are at similar altitude and geomagnetic latitude.

The Milagro cosmic-ray sky map~\citep{Abdo:2008kr} indicates two localized
regions of significant cosmic-ray excess called Regions A and B.  Region A is 
roughly elliptical with an angular size of about $15^\circ$, centered
at right ascension $\alpha=69.4^\circ$ and declination $\delta=13.8^\circ$.
Region B is larger, spanning a declination range $15^\circ < \delta < 50^\circ$
at right ascension $\alpha\simeq130^\circ$.  The relative intensity in Regions
A and B is $6\times 10^{-4}$ and $4\times 10^{-4}$, respectively.  Both regions
have also been observed with the Tibet AS$\gamma$ and ARGO-YBJ experiments.  In
a recent study based on $3.7\times 10^{11}$ events~\citep{ARGO-YBJ:2013gya},
the ARGO-YBJ experiment presented evidence for two additional excess regions
with lower relative intensity than Regions A and B.  Region 3 in the ARGO-YBJ
map is a rather elongated structure around $\alpha=240^\circ$, spanning the
declination range $15^\circ < \delta < 55^\circ$, with a maximum relative
intensity of $2.3\times 10^{-4}$.  Another new region (Region 4) around
$\alpha=210^\circ$ and $\delta=30^\circ$ has a maximum relative intensity of 
$1.6\times 10^{-4}$ and is currently the weakest statistically significant excess in the
ARGO-YBJ map.

In this paper, we present the results of a search for cosmic-ray anisotropy
in the northern sky with HAWC.  With current statistics, the HAWC cosmic-ray sky
map exhibits three regions of significantly enhanced flux.  The two strongest
excess regions coincide with Milagro Regions A and B (ARGO-YBJ Regions 1 and 2).
The location and relative intensity of the largest excess in Region A in HAWC and
Milagro data show some differences, but the excess observed with HAWC agrees
well with the excess seen by ARGO-YBJ.  A third significant excess in the HAWC
map coincides with Region 4 in the ARGO-YBJ map.  With HAWC, this region is
detected at almost twice the relative intensity observed by ARGO-YBJ, making it
more significant in HAWC despite the considerably shorter observation period.
A small excess in the HAWC map near Region 3 is currently not statistically
significant.

In addition to the arrival direction distribution, we also present the angular
power spectrum of the cosmic rays.  This analysis confirms the presence of a
strong dipole and quadrupole moment and shows significant power on angular
scales down to $12^\circ$ with current statistics.  The power spectrum can be
compared to the spectrum of the southern sky~\citep{Abbasi:2011ai} and to
recent model predictions that link the presence of higher order multipoles to
the dipole component using phase-space arguments~\citep{Ahlers:2013ima}.

The paper is organized as follows.  After a brief description of the HAWC detector 
(Section\,2), we describe the data set used in this analysis (Section\,3).
In Section\,4, we present the arrival direction distribution, an analysis of 
the excess regions, and the results of the angular power spectrum analysis.
The paper is summarized in Section\,5.

The paper focuses on the measurement of the small-scale anisotropy.  With less
than a full year of coverage, we expect that large-scale anisotropy
measurements are still contaminated by several effects that typically cancel
with one or more full years of continuous data, such as the dipole produced 
by the motion of the Earth around the Sun.  The measurement of the large-scale
anisotropy with HAWC will be the subject of a future publication.

\section{The HAWC Detector}
\label{sec:Detector}

The Earth's atmosphere is not transparent to cosmic rays and gamma rays at TeV 
energies.  The incoming primary particle interacts with a molecule in the atmosphere
and creates an extensive air shower, a huge cascade of secondary particles.
Ground-based detectors like HAWC need to reconstruct the properties of the
incoming cosmic rays from the particles of the shower cascade that reach the
observation level.  In HAWC, the secondary particles of the air shower cascade
are detected with instrumented water tanks, making use of the fact that the
relativistic particles of the shower cascade produce Cherenkov light when
traversing the water in the tanks.

The HAWC Observatory~\citep{Abeysekara:2013tka} is a 22,000~m$^2$ array of 
close-packed water Cherenkov detectors (WCDs).  Each WCD consists of a 
cylindrical steel water tank 4.5\,m in height and 7.3\,m in diameter.  A black plastic 
liner inside each tank contains 188,000~liters of purified water, and four 
photo-mulitplier tubes (PMTs) are attached to the liner on the floor of the tank: 
one central high-quantum efficiency 
Hamamatsu 10'' R7081 PMT and three Hamamatsu 8'' R5912 PMTs each at 1.8\,m 
from the center forming an equilateral triangle.  The PMTs face
upward to observe the Cherenkov light produced when charged particles from air
showers enter the tank.  When construction is complete, the observatory will 
comprise 300 water Cherenkov detectors with 1200 PMTs.

The signals from each PMT are transferred via RG59 coaxial cables to a counting
house in the center of the array where the pulses are amplified and shaped
using custom front-end electronics. The shaped pulse is compared with two
different discriminator levels, and the time over each level, high and low time
over threshold (ToT), are recorded as time-stamped edges with 100\,ps
resolution using CAEN VX1190A TDC modules. The shaping is such that ToT values
are proportional to the logarithm of the charge in the pulse.

Once the signals have been time-stamped the data are aggregated into an online
data stream.  Multiple online clients pull data from Single Board Computers
that poll the TDCs for data, combine the data into blocks, and buffer them for
readout.  The hits are sorted into an ordered time series, and a simple
multiplicity trigger is applied to identify candidate air shower events.  The
trigger condition requires at least 15 PMTs to be above threshold within a
sliding time window of 100\,ns. The triggered events are then written to disk.
As the data are being saved, an online reconstruction determines the arrival 
direction of the primary particle in real time.  However, since the detector 
has been growing and changing, in this analysis the angular reconstruction
was performed offline.

The hit times are calibrated to remove both a relative timing offset due to 
differences between individual PMT responses and a timing offset from 
the distribution of arrival times expected in an air 
shower~\citep{Abeysekara:2013tka}. The relative timing offset, or slewing offset, 
is the result of the combined response time of a specific PMT and the front-end 
electronics. The slewing offsets are determined with an on-site laser calibration 
system that sends pulses of varying intensities to each WCD while on-site 
computers record the PMT responses. After accounting for the relative timing 
difference between PMTs, the timing offset between a best-fit air shower 
front and the PMT hit times is calculated and subtracted in an iterative shower 
reconstruction procedure.
 
\begin{figure}[t]
  \centering
    \includegraphics[width=0.45\textwidth]{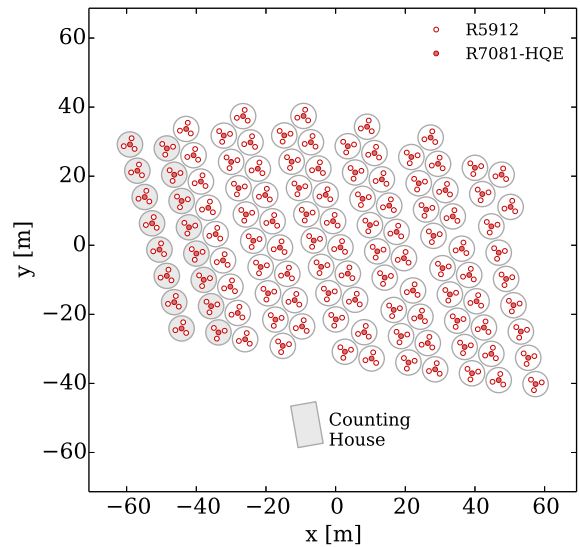}
    \caption{Layout of HAWC-95, with large shaded circles indicating the
     additional two rows of tanks present in HAWC-111.  The positions of the
     8'' R5912 PMTs are shown as small open circles and the 10'' R7081
     high-quantum efficiency PMTs are shown as small filled circles.}
    \label{fig:hawc-layout}
\end{figure}

\section{The Data Set}
\label{sec:DataSet}

The analysis in this paper uses data recorded between June 16, 2013, and
February 27, 2014.  Before August 12, 2013 the detector was operated with 95
tanks (HAWC-95), and afterwards with 111 tanks (HAWC-111). The layout of 
HAWC-95 and HAWC-111 is shown in Fig.\,\ref{fig:hawc-layout}.

For all analyses described in this paper, we apply additional cuts to improve 
the data quality.  To remove poorly reconstructed events, we require at least 30
triggered PMTs per event.  The remaining events have a median 
angular resolution of $1.2^\circ$ and a median energy of 2\,TeV according to the
detector simulation, which uses the CORSIKA code~\citep{Heck:1998vt} to
generate air showers in combination with a Geant4-based
simulation~\citep{Agostinelli:2002hh} of the detector response.  The angular
resolution for cosmic-ray showers derived from simulations agrees with the
estimate from the observation of the cosmic-ray shadow of the Moon, which is
described in this section.  The completed array is
expected to have an angular resolution of about $0.2^{\circ}$ for gamma rays
above 2\,TeV~\citep{Abeysekara:2013tza}. Gamma rays are present in this 
dataset but the gamma ray population is negligible with respect to cosmic rays. 
The significance of the gamma-ray background in this analysis will be the subject 
of future work.

\begin{figure}[t]
  \centering
    \includegraphics[width=0.45\textwidth]{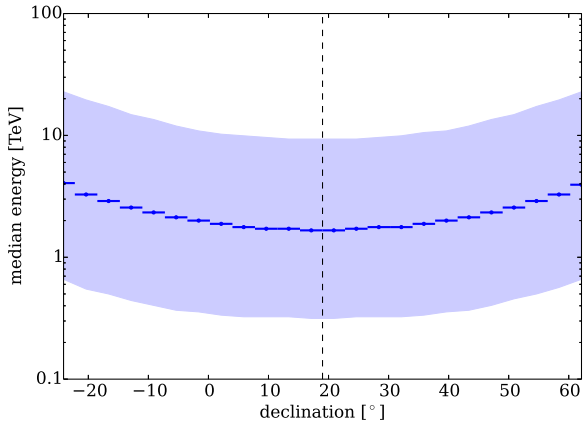}
    \caption{Median energy of the cosmic-ray flux observed in HAWC as a
     function of declination. The dashed line indicates the latitude of the
     detector, and the shaded region corresponds to a 68\% containing interval.}
    \label{fig:medianenergy}
\end{figure}

The median energy quoted above refers to the total cosmic-ray flux detected with
HAWC.  For localized excess regions, which are the main interest of this paper,
the median energy is a function of the region's declination.  The dependence
of the median energy on declination is shown in Fig.\,\ref{fig:medianenergy}, 
together with the central 68\% quantile.  The median cosmic-ray energy ranges 
from 1.7\,TeV at a declination of $19^\circ$ to 4\,TeV at declinations below 
$-20^\circ$ and above $60^\circ$, near the border of the field of view.

The data are further reduced by requiring that only full and continuous
sidereal days of data runs be used.  This produces a nearly uniform exposure
across right ascension, providing for an easy interpretation of the
significance map since exposure changes only in declination.  The only
non-uniformity in right ascension is the result of diurnal variations in the
cosmic-ray rate.  This has been confirmed using a set of simulated data with an
event time distribution that matches the actual distribution.

During most of HAWC-95, construction took priority over data-taking and, while
data-taking is largely uninterrupted in the HAWC-111 data set, many HAWC-111
data runs do not last a full sidereal day due to detector upgrades and tests of
the data acquisition and calibration systems.  As the procedures for upgrades
and calibration of the detector are improved and construction ceases, the
number of full sidereal days will approach the number of days of detector
up-time.

The resulting data set contains $113$ full and continuous sidereal days during
which the detector collected $4.9\times10^{10}$ well-reconstructed events - about
1.5 times the number of events in the IceCube 2009-2010 anisotropy
data set~\citep{Abbasi:2011ai}, but still four times smaller than the total
number of Milagro events~\citep{Abdo:2008kr} and almost 8 times smaller than
the number of ARGO-YBJ events~\citep{ARGO-YBJ:2013gya}.

\begin{figure}[t]
  \centering
    \includegraphics[width=0.45\textwidth]{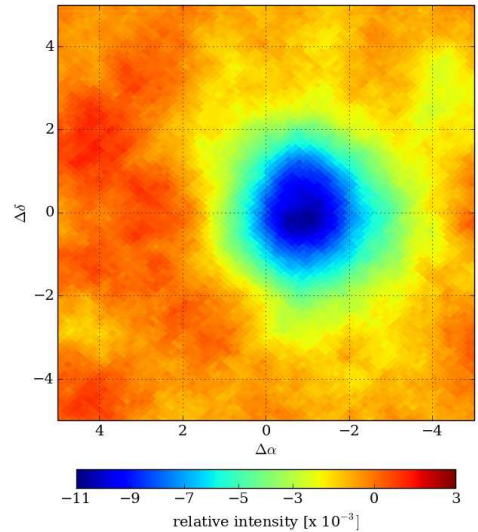}
    \caption{Relative intensity of the cosmic-ray flux in a sky map centered on 
     the position of the Moon.  $\Delta\alpha$ and $\Delta\delta$ are the right 
     ascension and declination of the cosmic rays with respect to the right 
     ascension and declination of the Moon.  The Moon shadow is shown for 
     113 days of HAWC-95/111 data.  The deficit corresponds to a significance of
     $24\,\sigma$.}
    \label{fig:moon}
\end{figure}

The angular resolution for cosmic rays and the energy of the isotropic
cosmic-ray flux triggering HAWC can be verified by studying the
cosmic-ray shadow of the Moon.  The Moon produces an observable deficit
in the nearly isotropic flux of cosmic-ray air showers incident at Earth,
and the width and shape of the deficit indicate the instrument's
point-spread function for cosmic rays.  The apparent position of the Moon 
shadow differs from its true position because of deflections of the cosmic 
rays in the geomagnetic field.  From simulations described in detail 
in~\citet{Abeysekara:2013qka}, the geomagnetic deflection $\delta\theta$
of particles arriving at the HAWC altitude and geographic location can 
approximately be summarized by
\begin{linenomath}
\begin{equation}
  \delta\theta\simeq 1.6^\circ\cdot Z\left(\frac{E}{\text{TeV}}\right)^{-1}~,
\end{equation}
\end{linenomath}
where $E$ and $Z$ are the cosmic-ray energy and charge, respectively.

HAWC-111 observes the shadow of the Moon at a significance of $13\,\sigma$ 
per month.  To study the location and shape of the deficit, we 
produce sky maps centered on the position of the Moon.  The relative intensity 
of the cosmic-ray flux as a function of the relative right ascension $\Delta\alpha$ 
and declination $\Delta\delta$ is shown in Fig.\,\ref{fig:moon} and was obtained by 
subtracting the calculated equatorial coordinates of the Moon 
$(\alpha_\text{Moon},\delta_\text{Moon})$ from the right ascension $\alpha$ 
and declination $\delta$ of each reconstructed cosmic-ray shower.  
In 113 days of HAWC-95/111 data, the statistical significance of the deficit
in the cosmic-ray flux is about $24\,\sigma$.

From a fit of a two-dimensional Gaussian to the deficit 
region, we find that the observed Moon shadow is offset by 
$-1.05^\circ\pm 0.05^\circ$ in $\Delta\alpha$ and $-0.02^\circ\pm 0.06^\circ$ 
in $\Delta\delta$ and has a width of $1.26^\circ\pm 0.05^\circ$.  According to 
simulations, the cosmic-ray deflection in the magnetic field also leads to a 
slight broadening of the width of the Moon shadow, so it should
be interpreted as an upper limit on the angular resolution of 
the detector.  The shift of the shadow indicates that the observed energies 
are dominated by proton-initiated showers of about 1.6\,TeV.  Both angular 
resolution and energy scale are consistent with the predictions from simulations.

\section{Analysis}
\label{sec:analysis}

\subsection{Relative Intensity and Significance Map}
\label{subsec:map}

The search for anisotropy is based on techniques described
in~\citet{Abdo:2008kr} and~\citet{Abbasi:2011ai}.  To produce a sky map of the
relative intensity of the cosmic-ray flux requires a comparison of the data
to a reference map which represents the response of the detector to an
isotropic cosmic-ray flux.  This reference map is not itself isotropic, as
atmospheric effects cause diurnal changes in the cosmic-ray rate and the
asymmetric shape of the HAWC-95/HAWC-111 tank configuration leads to an uneven
event distribution in right ascension.  Because the reference map needs to
account for these and other effects, which are difficult or impossible to
simulate at the required level of accuracy, it has to be constructed from the
data themselves.

We begin by binning the sky into an equal-area grid in equatorial coordinates
with an average pixel size of $0.23^\circ$ using the HEALPix
library~\citep{Gorski:2004by}.  The resolution of the HEALPix pixelation of the
sphere is defined by a parameter $N_{\mathrm{side}}$ which is related to the
number of pixels by $N_{\mathrm{pix}}=12\,N_{\mathrm{side}}^{2}$.  In this
analysis, we chose $N_{\mathrm{side}}=256$, so the sky is originally divided
into 786\,432 pixels\footnote[1]{For the analysis of the Moon shadow in
Section\,\ref{sec:DataSet}, $N_{\mathrm{side}}=512$ was used.}.  Since HAWC 
covers the sky at declinations between $-26^\circ$ and $64^\circ$, the total 
number of independent pixels is 525\,716.

A binned data map $N(\alpha,\delta)$ is used to store the arrival directions of
air showers reconstructed from data.  The reference map $\langle
N(\alpha,\delta)\rangle$ is produced using the direct integration technique
described in~\citet{Atkins:2003ep}, adapted for the HEALPix grid.  We begin by
collecting all events recorded during a predefined time period $\Delta t$ and
convolve the local arrival direction distribution with the detector event
rate.  The method effectively smooths out the true arrival direction
distribution in right ascension on angular scales of roughly $\Delta
t\cdot15^\circ~\text{hour}^{-1}$, so the analysis is only sensitive to
structures smaller than this characteristic angular scale.  The direct
integration method produces a reference map with the same underlying local
arrival direction distribution and the same event rate as the data.  Therefore,
any effects from temporal variations in the cosmic-ray rate or from the detector
geometry appear in the data map as well as in the reference map and cancel when
the two are compared to produce maps of significance or relative intensity.
The relative intensity map is calculated as 
\begin{linenomath}
\begin{equation}
  \delta I(\alpha_i,\delta_i)
    = \frac{\Delta N_i}{\langle N\rangle_i}
    = \frac{N(\alpha_i,\delta_i) - \langle N(\alpha_i,\delta_i)\rangle}
           {\langle N(\alpha_i,\delta_i)\rangle},
\end{equation}
\end{linenomath}
which gives the amplitude of deviations from the isotropic expectation in each
angular bin $i$.  

We emphasize that this algorithm estimates the reference level by averaging
the number of events over a fixed declination band.  Because different
declination bands have different normalizations, the method is not sensitive
to anisotropy that depends only on declination, {\it i.e.}, with constant
relative intensity in right ascension.  Studies using simulated
data~\citep{santander:2013} show that this does not affect typical small-scale
structure, but it reduces the sensitivity of the method to large-scale structure.
As an example, for a pure dipole tilted at some angle with respect to the
equatorial plane, the method is only sensitive to the {\it projection} of the
dipole onto the equatorial plane.

To improve the sensitivity to features on angular scales larger than the pixel
size, we apply a smoothing procedure which takes the event counts in each pixel
and adds the counts from neighboring pixels within a radius $\theta$.  This is
equivalent to convolving the map with a top hat function of radius $\theta$.
Applied to both the data map and the reference map, the procedure leads to maps
with the original binning, but neighboring pixels are no longer independent and
pixel values become highly correlated over a range $\theta$.  In this paper, we
use $\theta=10^\circ$, the same scale used in~\citet{Abdo:2008kr}.  This scale
is a compromise, since the optimal $\theta$ varies from region to region, with
no single scale appropriate for the entire sky map.  However, $\theta=10^\circ$
displays all the relevant features and allows us to analyze the shape of the
anisotropy.

Gamma rays are present in the data set, but since the analysis is not optimized 
for gamma rays and we apply a smoothing radius of 10$^\circ$, far larger than 
the optimal smoothing radius for point sources, even the brightest TeV 
gamma-ray sources, such as the Crab, do not appear as a significant excess in 
the smoothed maps.

The significance of the deviation of the data from the isotropic expectation in
each bin is calculated using the method described in~\citet{Li:1983fv}.  In
this method, the statistical uncertainty of the number of events in each bin of
the reference map depends on the quantity $\alpha_{\text{Li-Ma}}$, the ratio of
time spent on-source to time spent off-source.  The effective value for
$\alpha_{\text{Li-Ma}}$ depends on the integration time $\Delta t$, smoothing
radius $\theta$, and declination $\delta$ and is calculated using
\begin{linenomath}
\begin{equation}
 \alpha_{\text{Li-Ma}}
    = \frac{\pi\,\theta^2}{2\,\theta\,(15^\circ/\text{hr})~\Delta t~\cos\delta}~.
\end{equation}
\end{linenomath}
For $\Delta t=24$\,h, $\theta=10^\circ$, and $\delta=0^\circ$, the value for
$\alpha_{\text{Li-Ma}}$ is 0.0436.  

Direct integration requires the local arrival direction distribution and thus
the acceptance of the detector to be stable throughout the time period $\Delta
t$.  Using a $\chi^2$-difference test to compare local arrival direction
distributions over various time periods, we find that the shape of these
distributions is stable over very long periods (up to several weeks) and
changes only when the detector geometry changes (for example at the time of the 
switch from HAWC-95 to HAWC-111).  The high stability of the detector
allows us to use $\Delta t=24$\,h in this analysis.  As described above, this
value preserves features on all angular scales, including the dipole moment.
The angular power spectrum can be determined directly from this relative
intensity map (see Section\,\ref{subsec:powerspectrum}).

\begin{figure*}[t]
  \centering
    \includegraphics[width=0.7\textwidth]{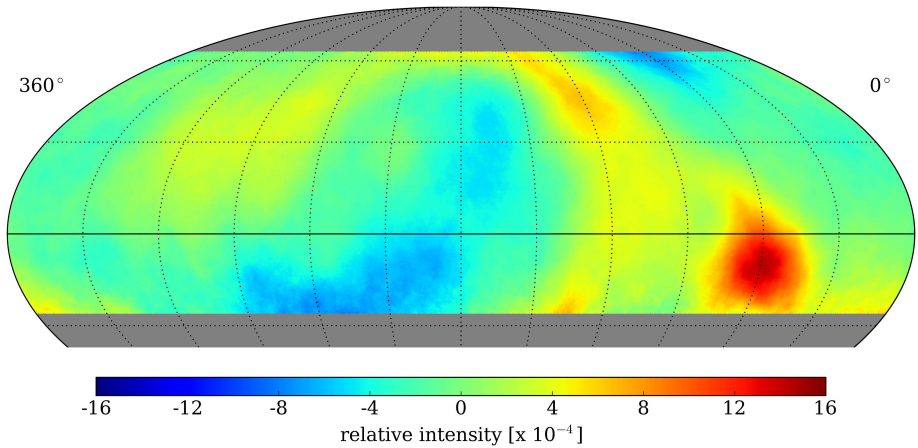}
    \caption{Relative intensity of the cosmic-ray flux for 113 days of 
     HAWC-95/111, in equatorial coordinates.  Right ascension runs from $0^\circ$     
     to $360^\circ$ from right to left.  The solid horizontal line denotes a 
     declination of $0^\circ$.  Lines of equal right ascension and declination 
     are separated by $30^\circ$.  The map contains $4.9\times 10^{10}$ events.  
     An integration time of $\Delta t=24$\,h is used to access the largest features 
     present in the map.  The map is shown with $10^{\circ}$ smoothing applied.}
    \label{fig:skymap_24h_relint}
\end{figure*}

The relative intensity of the cosmic-ray flux for an integration time of $\Delta t=24$\,h 
and a smoothing scale $\theta=10^\circ$ is shown in Fig.\,\ref{fig:skymap_24h_relint}.  
Several significant features appear in this map.  The
localized excess region at right ascension $60^\circ$ and declination
$-10^\circ$, which roughly coincides with Region A of the Milagro map and (more
accurately) with Region 1 of the ARGO-YBJ map, dominates the sky map.  In
addition, the large-scale structure of the cosmic-ray flux, with its broad
deficit region at $200^\circ$, is clearly visible in this map.  The large-scale
structure potentially distorts any smaller structures, enhancing their excess
in the region near the maximum of the large-scale structure and suppressing
them near the broad minimum.  As we are interested in structure on scales
smaller than $60^\circ$, corresponding to multipoles $\ell>3$, we need to
remove the lower order multipoles from the sky map.  We apply two different
methods to remove or suppress the $\ell\leq3$ term.

In the first method, we directly fit the relative intensity map to the sum
of the monopole ($\ell=0$), dipole ($\ell=1$), quadrupole ($\ell=2$), and 
octupole ($\ell=3$) terms of an expansion in Laplace spherical harmonics
$Y_{\ell m}$.  The fit function $F(\alpha, \delta)$ therefore has the form
\begin{linenomath}
\begin{equation}
F(\alpha_i,\delta_i) =
    \sum_{\ell=0}^{3}
    \sum_{m=-\ell}^{\ell} a_{\ell m} Y_{\ell m}(\pi-\delta_i,\alpha_i)~~,
\end{equation}
\end{linenomath}
where $(\alpha_i,\delta_i)$ are the right ascension and declination of the
$i^{\mathrm{th}}$ pixel and the $a_{\ell m}$ are the 16 free parameters of the fit.  
We then subtract the fit result from the map, and analyze the residual map.  

\begin{figure*}[t]
  \centering
    \includegraphics[width=0.7\textwidth]{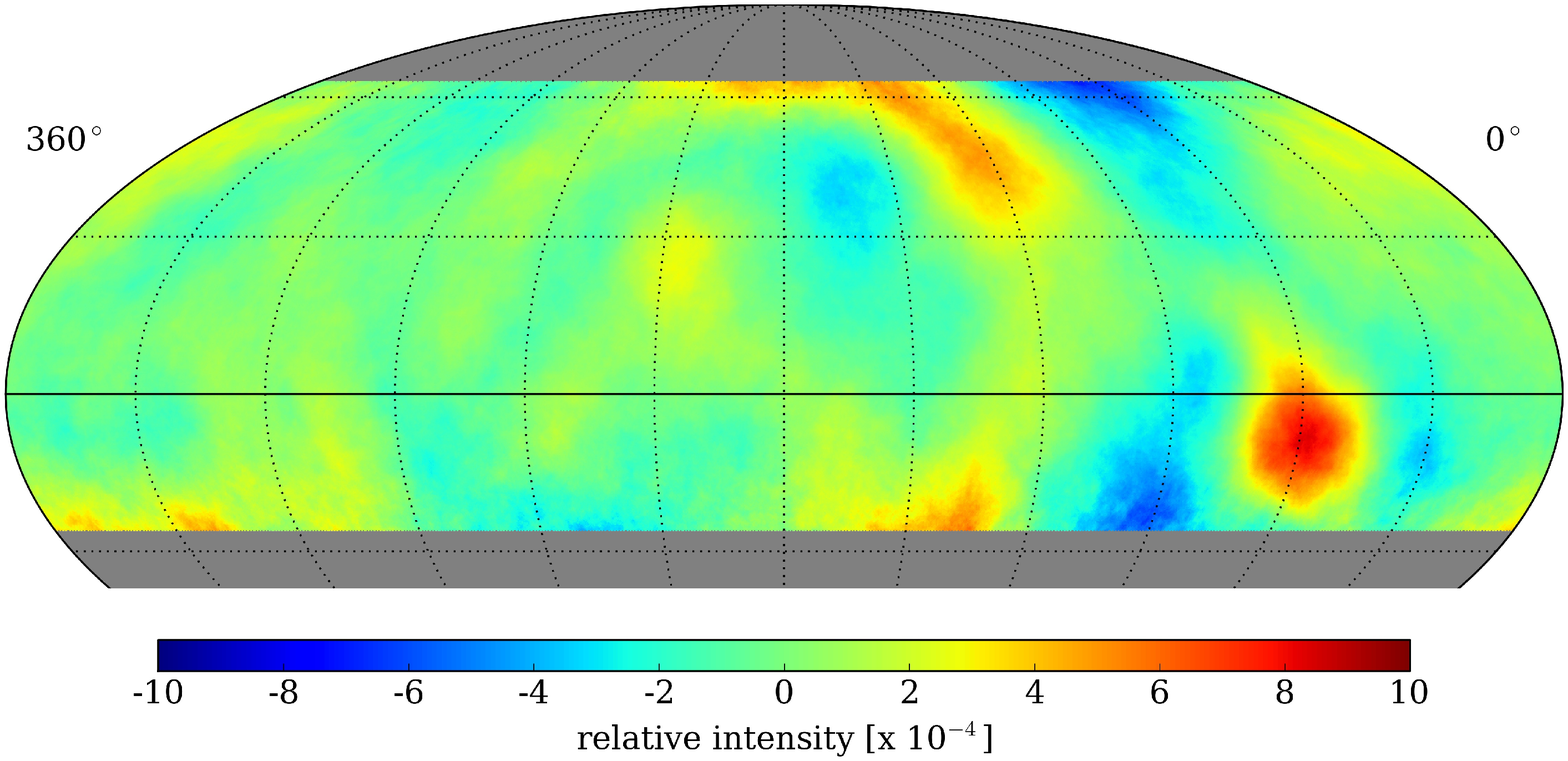}
    \includegraphics[width=0.7\textwidth]{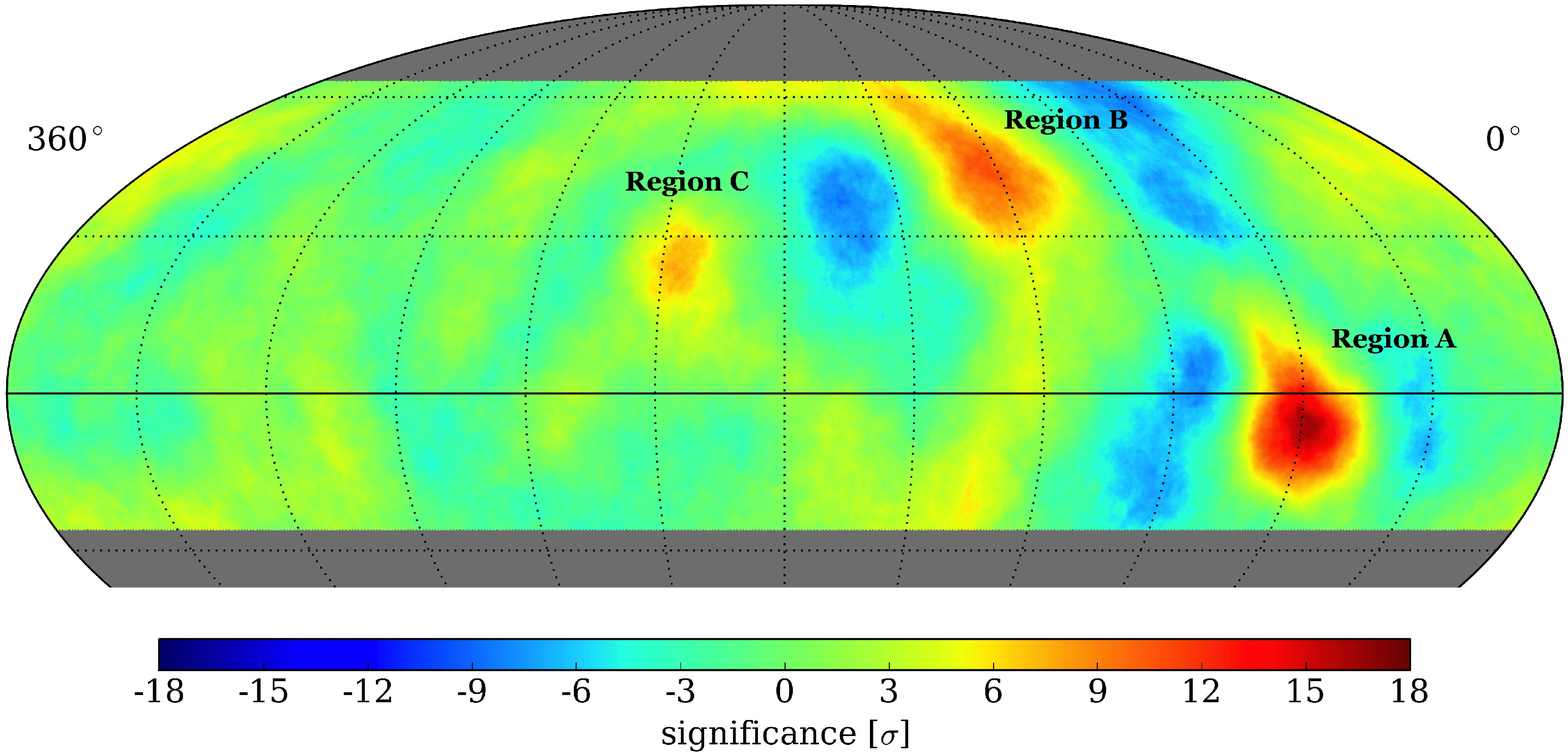}
    \caption{Relative intensity {\slshape (top)} and pre-trial significance 
     {\slshape (bottom)} of the cosmic-ray flux after fit and subtraction 
     of the dipole, quadrupole, and octupole term from the map shown in 
     Fig.\,\ref{fig:skymap_24h_relint}.  The map is shown with $10^{\circ}$ 
     smoothing applied. 
    }
    \label{fig:skymap_24h_multipolesub}
\end{figure*}

We perform the fit on the 525\,716 pixels of the 
relative intensity map that lie in the field of view of HAWC.  
The $\chi^2/\mathrm{ndf}=527\,282/525\,700$ corresponds to a 
$\chi^2$-probability of 6.0\%.  The marginal probability indicates that 
additional smaller structure is still present in the data.  Note that this fit 
gives a significantly better result than the fit with $\ell_\text{max}=2$ only 
(DC offset + dipole + quadrupole), corresponding to a $\chi^2$-difference of $262$
with 7 degrees of freedom.  The residual map in relative intensity (top) and 
significance (bottom) are shown in Fig.\,\ref{fig:skymap_24h_multipolesub}. 

\begin{figure*}[t]
  \centering
  \includegraphics[width=0.7\textwidth]{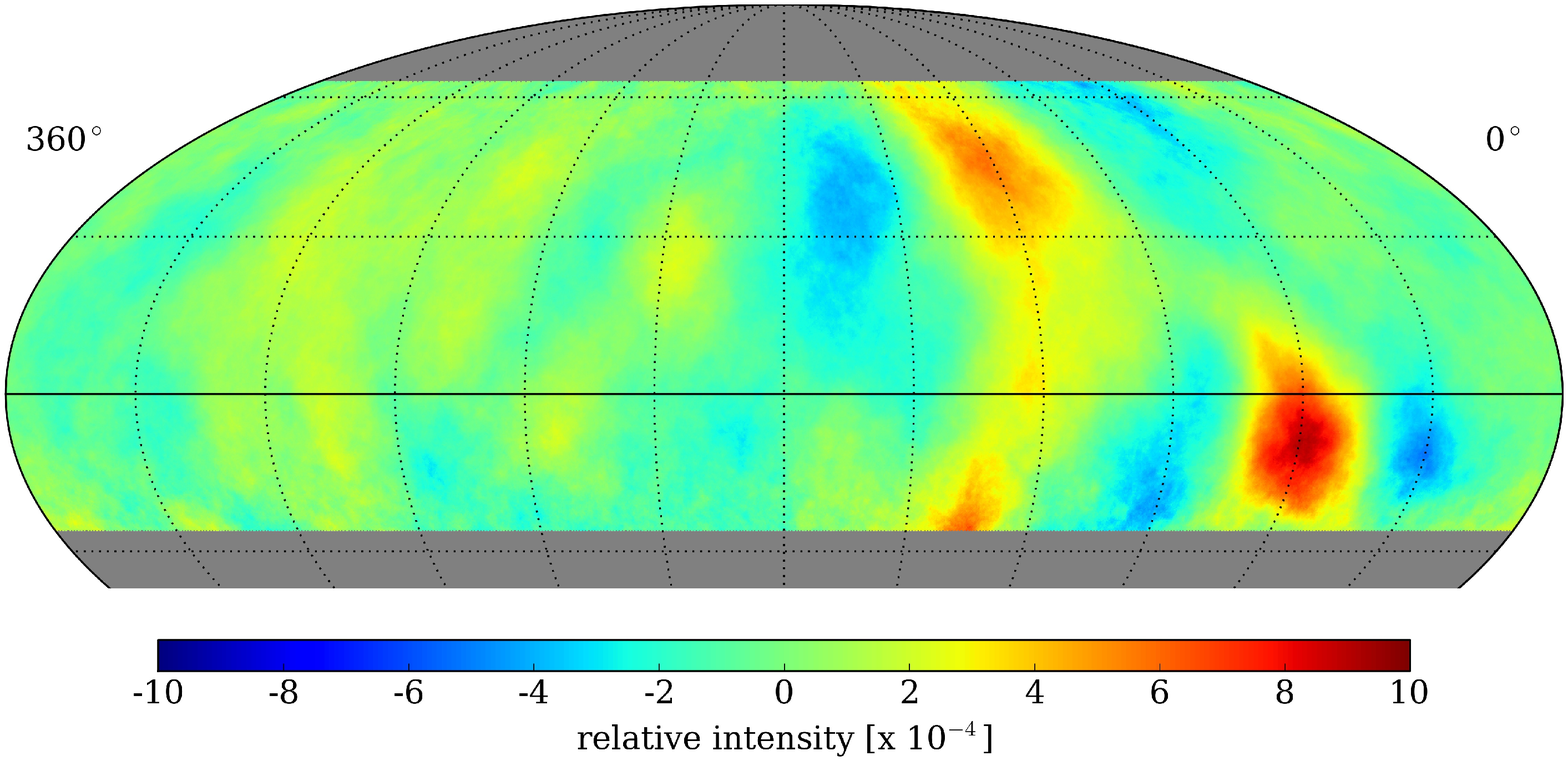}
  \includegraphics[width=0.7\textwidth]{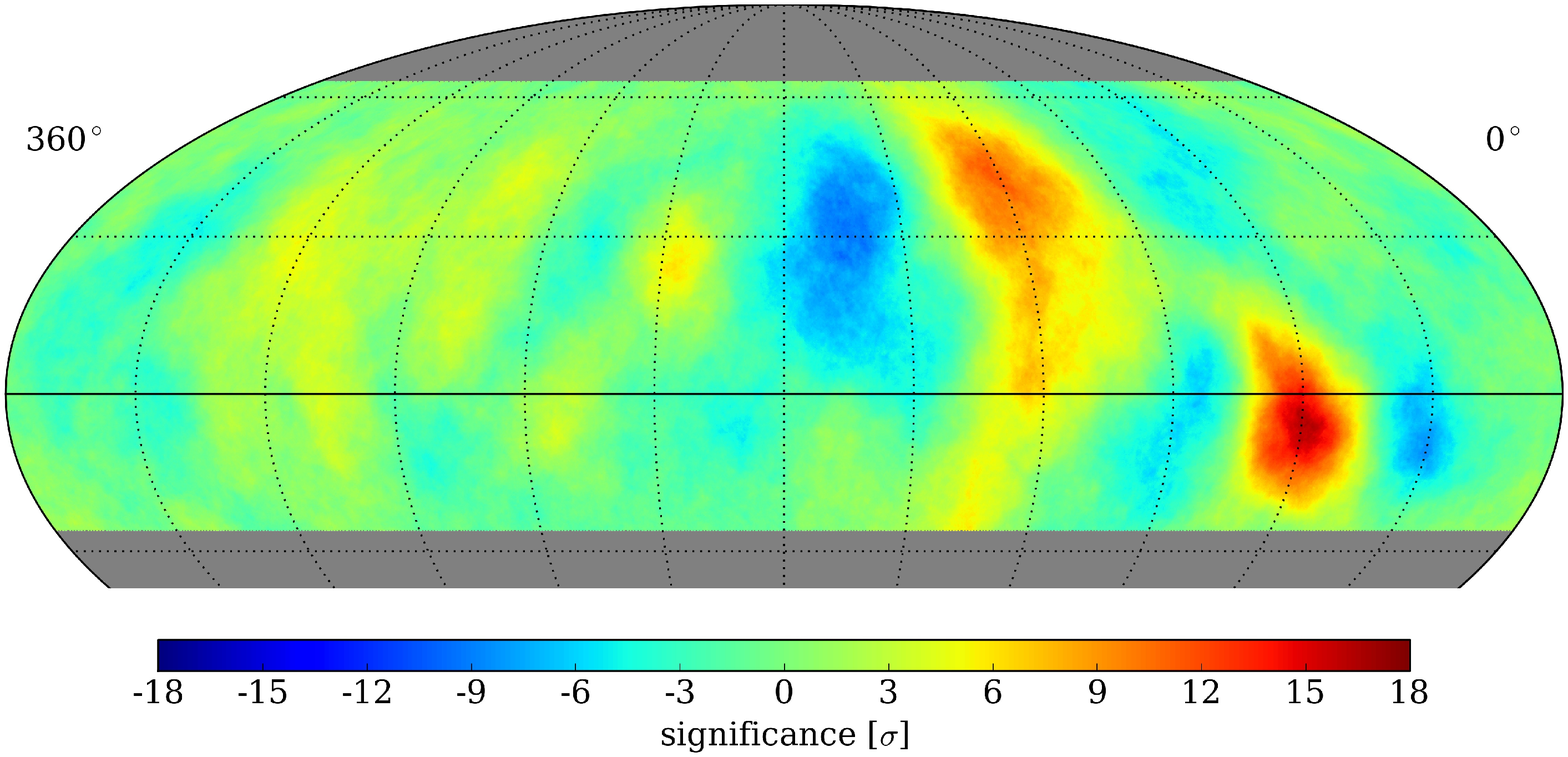}
    \caption{Relative intensity {\slshape (top)} and pre-trial significance {\slshape (bottom)} 
     of the cosmic-ray flux using a background estimated from direct integration with 
     a time period $\Delta t=4$\,h.  The map is shown with $10^{\circ}$ smoothing applied.}
    \label{fig:skymap_4h}
\end{figure*}

The second method uses a shorter integration time, $\Delta t=4$\,h, to filter
any structure with angular extent greater than $60^\circ$.  In
Fig.\,\ref{fig:skymap_4h}, we show the relative intensity (top) and significance
maps (bottom) produced with this method.  A comparison between 
Fig.\,\ref{fig:skymap_24h_multipolesub} and
Fig.\,\ref{fig:skymap_4h} shows that the maps are largely equivalent.  While
regions A and C agree well in shape and relative intensity, region B extends
into mid-latitudes for the $\Delta t=4$\,h map.

There are also regions of strong deficits visible, typically on both sides of
the strong excess regions.  The appearance of these deficit regions, correlated
with the excess regions, is a well-known artifact of the method~\citep{Abdo:2008kr}.  
They appear because the background near strong excesses is overestimated due to the 
fact that the excess events are part of the background estimation.

The two methods to remove the large-scale anisotropy are affected by different 
systematic uncertainties.  Estimating the background using $\Delta t=24$\,h and 
explicitly subtracting lower order multipoles should, in principle, minimize 
artifacts from the presence of strong excesses described above.  However, because 
of the incomplete sky coverage, the removal of the lower order multipoles can 
potentially affect higher order terms, too.  This effect is studied with the 
angular power spectrum analysis described in Section\,\ref{subsec:powerspectrum} 
and is found to be small in HAWC data.  Filtering the low order multipoles by 
choosing a short integration time $\Delta t$ also influences higher order 
multipoles (in a less transparent way than the direct subtraction), and it 
depends on the choice of $\Delta t$.  

In the following analysis, we estimate the systematic error on the relative 
intensity of cosmic-ray excess regions by comparing the intensity obtained with 
the two methods, and, in addition, by comparing two different integration times 
(3\,h and 4\,h) which are both found to preserve the power in the higher order 
multipoles of the angular power spectrum (Section\,\ref{subsec:powerspectrum}). 
The larger difference of the two alternative methods is taken as the systematic 
uncertainty reported in Section\,\ref{subsec:results} for the various regions of 
excess.  The cosmic-ray dipole caused by the motion of the Earth around the Sun 
can potentially distort any sidereal large-scale structure, but should have no 
effect on the small-scale structure.

\subsection{Results and Discussion}
\label{subsec:results}

\begin{figure*}[t]
  \centering
  \includegraphics[width=0.3175\textwidth]{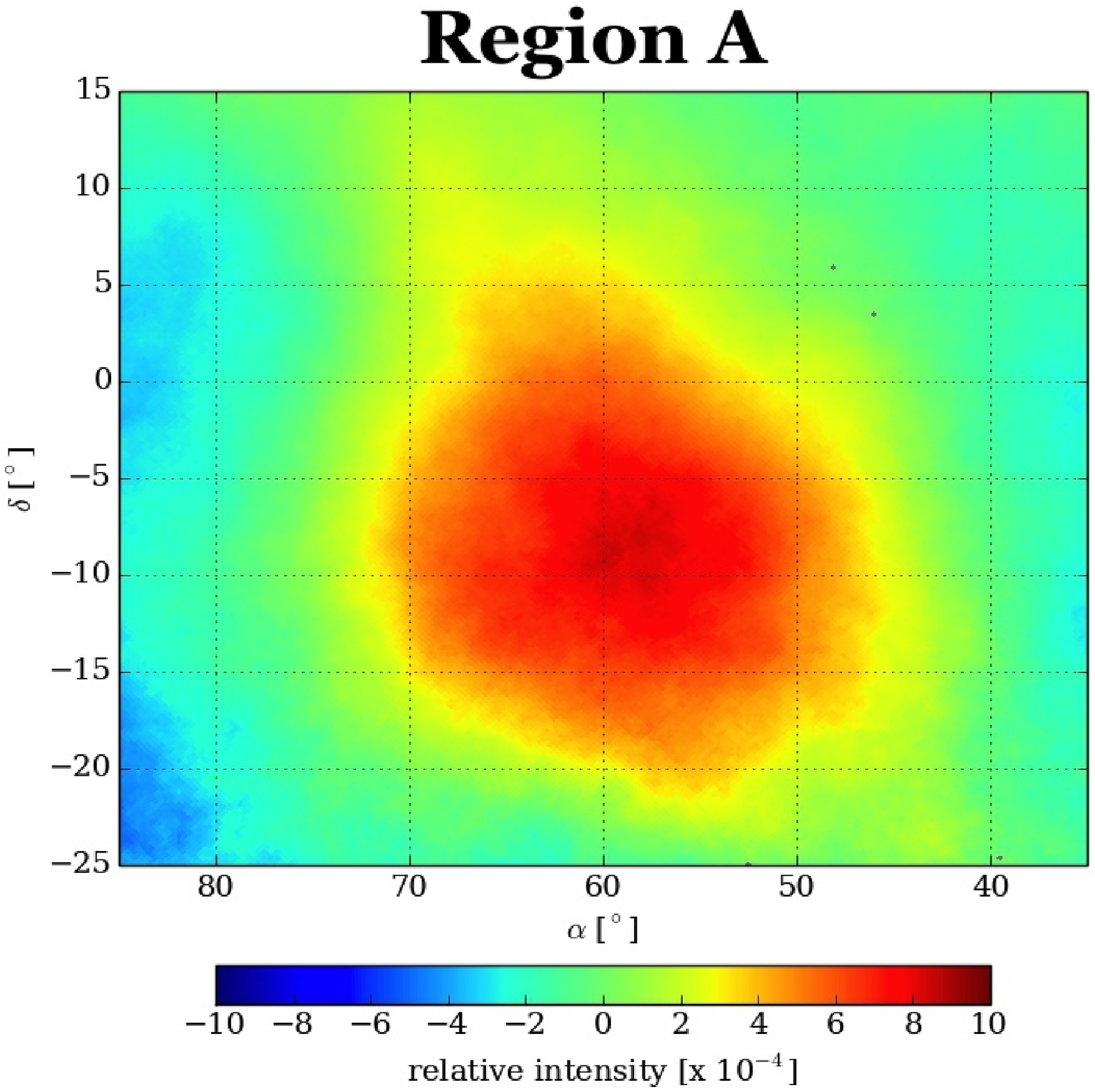}
  \includegraphics[width=0.3175\textwidth]{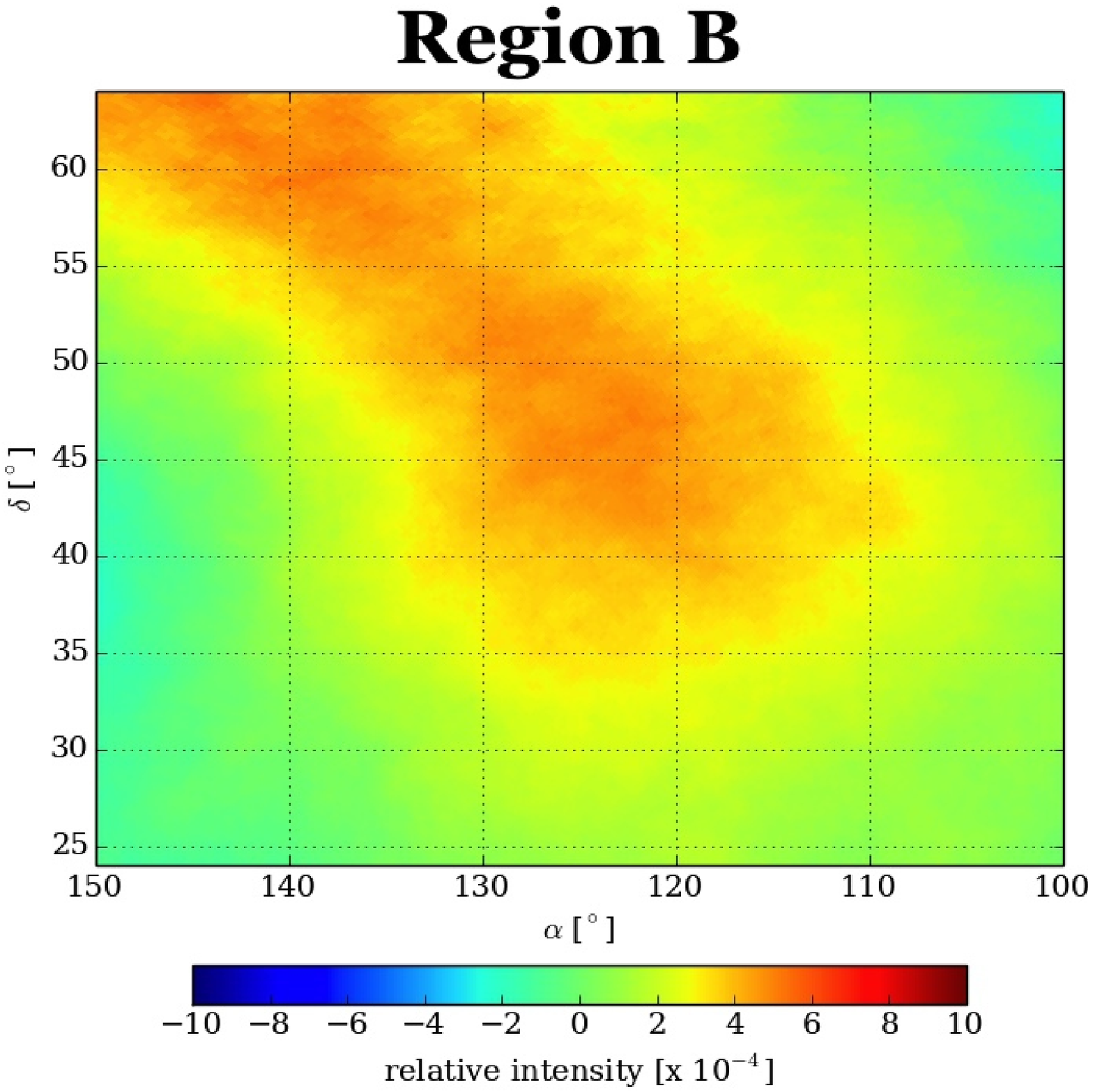}
  \includegraphics[width=0.3175\textwidth]{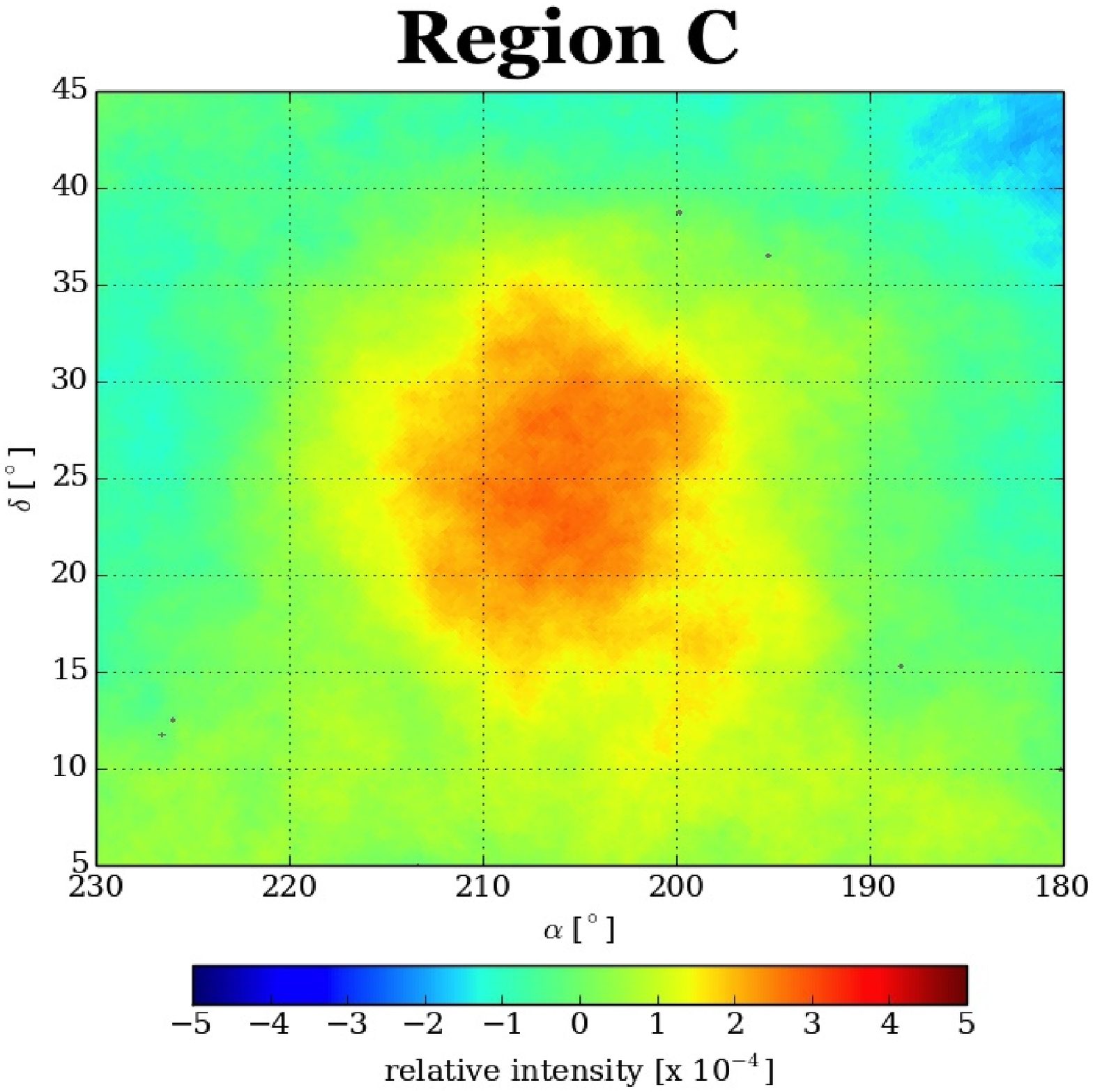}
  \includegraphics[width=0.3175\textwidth]{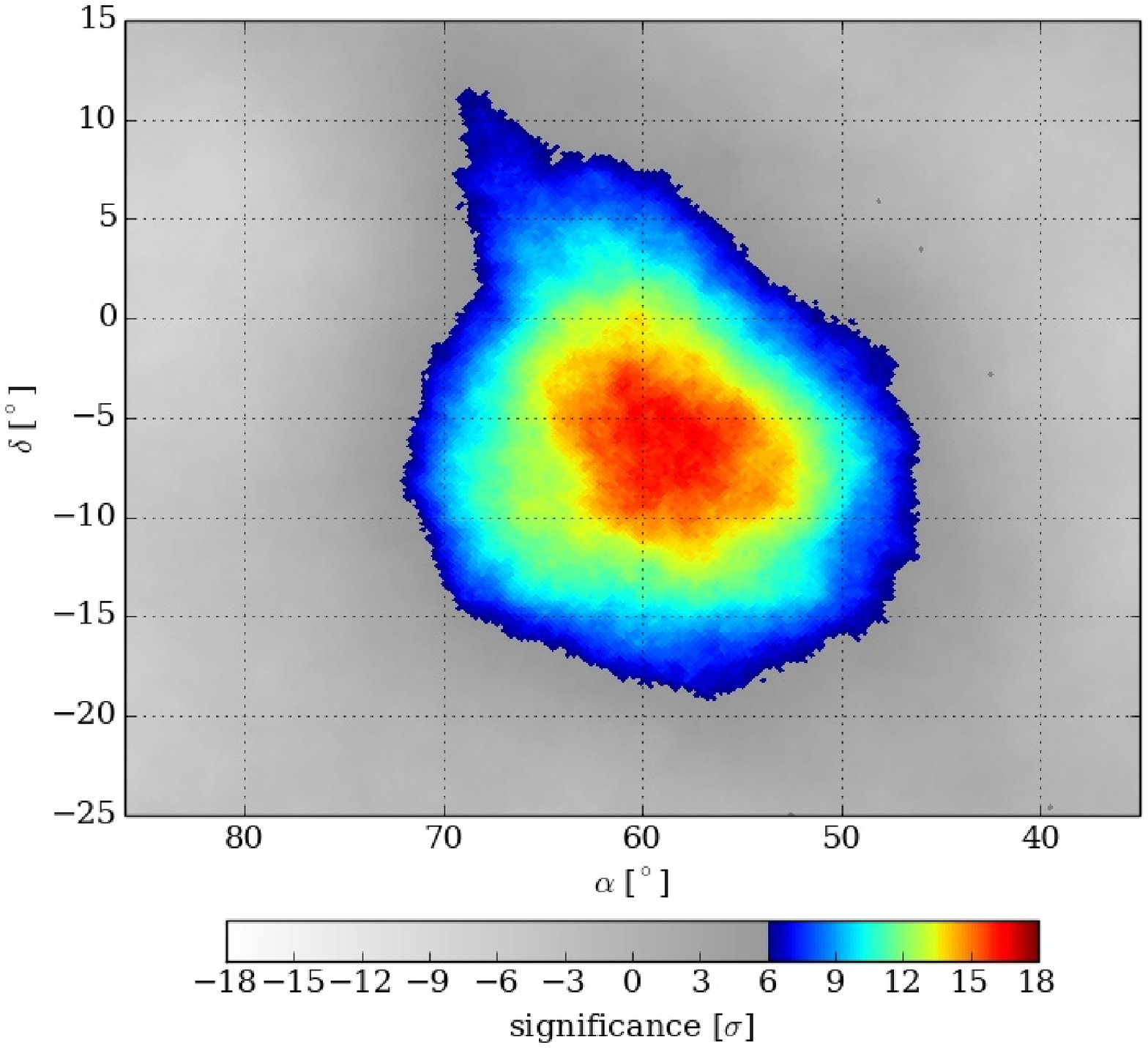}
  \includegraphics[width=0.3175\textwidth]{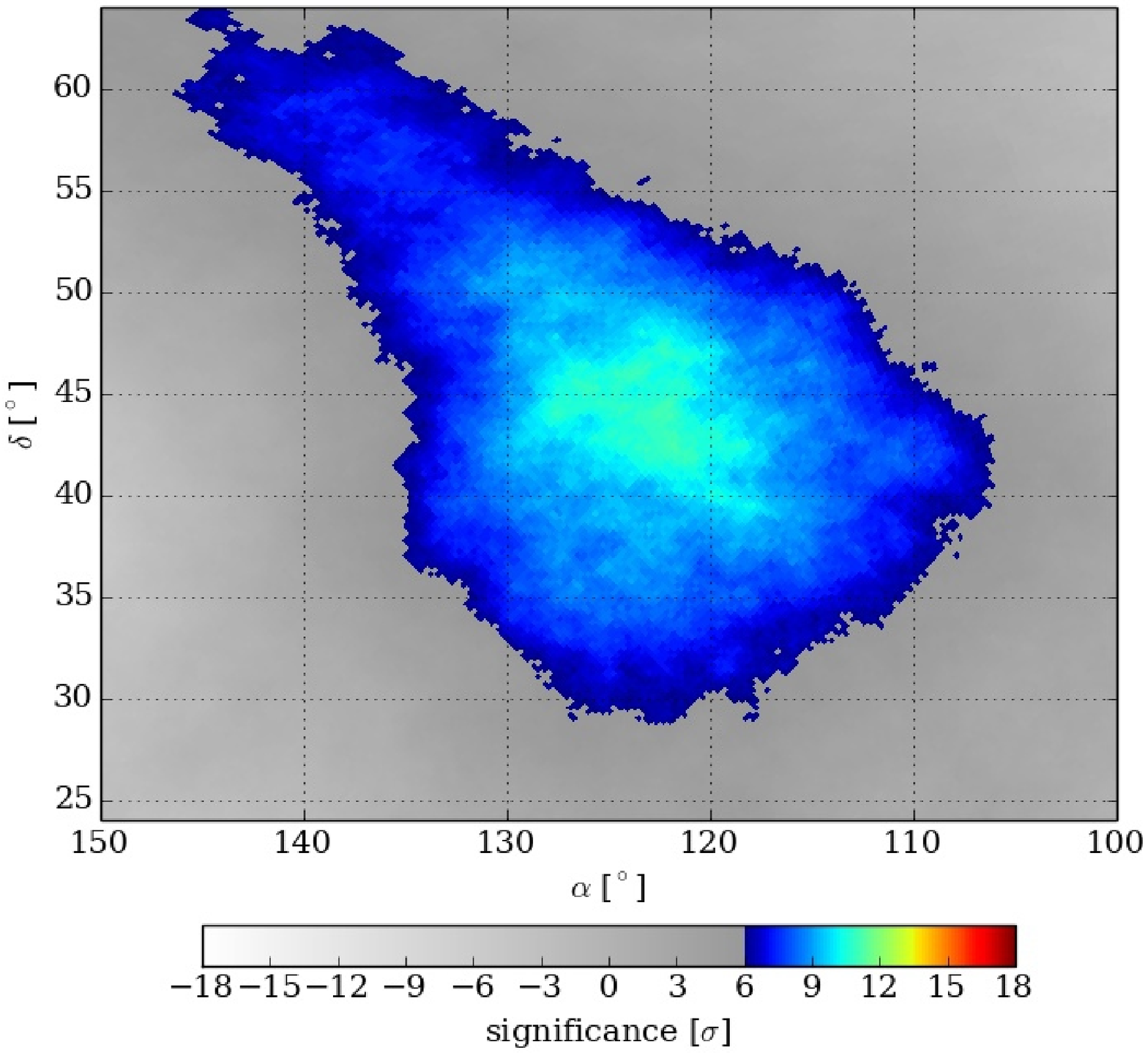}
  \includegraphics[width=0.3175\textwidth]{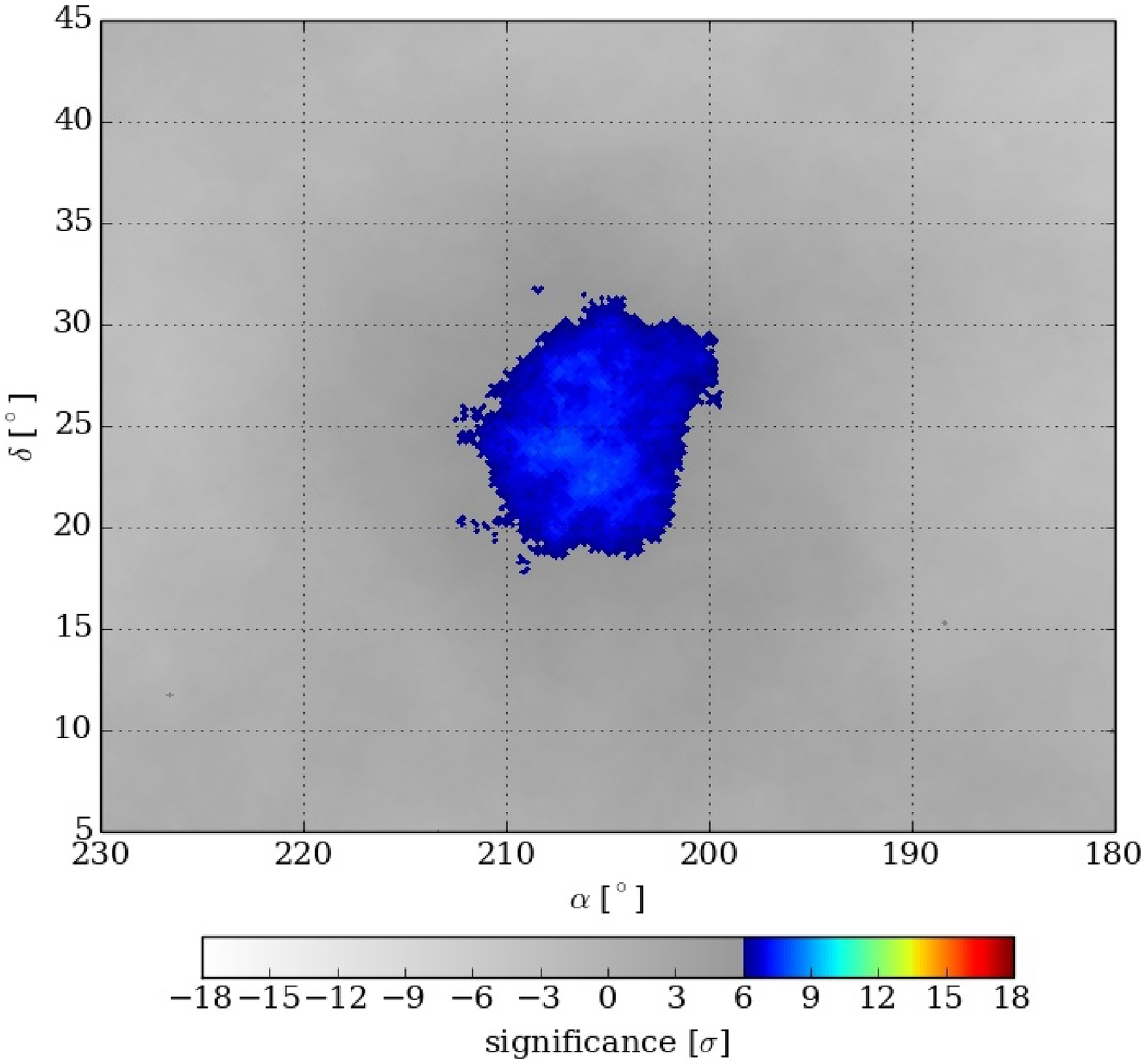}
  \caption{Relative intensity {\slshape (top row)} and pre-trial significance 
     {\slshape (bottom row)} of the cosmic-ray flux in the vicinity of Region A 
     {\slshape (left)}, Region B {\slshape (center)}, and Region C {\slshape (right)},
     from the map shown in Fig.\,\ref{fig:skymap_24h_multipolesub}.
  }
  \label{fig:regionABC}
\end{figure*}

After the elimination of the large-scale structure, the residual HAWC
cosmic-ray sky map shows several prominent features, notably three regions of
excess flux with high significance.  The strongest excess, with a pre-trial
significance of $17.0\,\sigma$, is found at $\alpha=57.5^\circ$ and
$\delta=-6.3^\circ$ and corresponds to Region A in the Milagro sky map and
Region 1 in the ARGO-YBJ sky map.  The relative intensity of the excess in this
region peaks at $(8.5\pm 0.6 \pm 0.8)\times 10^{-4}$, where the first error is
statistical and the second error is systematic.  A detailed map of the
morphology of this region is shown in the left panel of
Fig.\,\ref{fig:regionABC}.  The median cosmic-ray energy at this declination is
2.1\,TeV.  For comparison, we also fit a two-dimensional Gaussian function to
the relative intensity map around Region A.  The center is located at
$\alpha=60.0^\circ\pm 0.7^\circ$ and $\delta=-7.1^\circ\pm 0.8^\circ$, with an
amplitude of $(10.1\pm 1.2)\times 10^{-4}$.  The width is $7.1^\circ\pm
1.3^\circ$ in right ascension and $7.8^\circ\pm 1.3^\circ$ in declination.

The location and relative intensity of Region A in the HAWC sky map are
consistent with the ARGO-YBJ measurement, but there are notable differences
compared to Milagro.  The peak relative intensity in HAWC is a factor of 1.5
higher than in Milagro, but the locations of the peaks also differ.  While the
HAWC excess extends up to $\delta=15^\circ$, the most significant peak is
observed in the Southern Hemisphere at $\delta=-6.3^\circ$, at the edge of the
field of view of Milagro.  At the location of the centroid of Milagro's Region
A, the relative intensity in HAWC is only $(1.5\pm 0.4)\times 10^{-4}$, a
factor of 4 smaller than the Milagro excess.  A possible reason for this
discrepancy is that the median energy of the Milagro data is higher than in
HAWC and that the upper part of Region A, where Milagro observes the largest
excess, is brighter at higher energies.  The energy dependence of the Region A
excess will be studied in more detail in Section\,\ref{subsec:excess}.

Because of its more southerly latitude ($19^\circ$N) compared to Milagro 
($36^\circ$N) and ARGO-YBJ ($30^\circ$N), HAWC observes the lower part 
of Region A at a more favorable zenith angle.  This can account for the 
fact that the significance of Region A in HAWC is already as strong as 
in Milagro even though the HAWC data set is still considerably smaller.

The elongated excess around $\alpha=120^\circ$, identified as Region B in the
Milagro map and Region 2 in the ARGO-YBJ map, extends over a wide range of
declinations.  It is most significant at $(122.1^\circ, 43.8^\circ)$ with a pre-trial
significance of $11.2\,\sigma$ and a relative intensity of $(5.2 \pm 0.6 \pm
0.7)\times 10^{-4}$.  The morphology of this region is shown in the center
panel of Fig.\,\ref{fig:regionABC}.

There is considerable uncertainty in the shape of Region B.  It appears as two
separate regions in Fig.\,\ref{fig:skymap_24h_multipolesub}, one at high
northern latitude and one at declination $\delta<0^\circ$.  The two regions are
connected by band-like structure with lower relative intensity.  The map
produced with an integration time of $\Delta t=4$\,h (Fig.\,\ref{fig:skymap_4h})
also shows these regions, but the shape of the upper
region is broader and the intensity of the connecting band is brighter.  Region
B essentially spans almost the entire declination range visible to HAWC.  It is
also the only small-scale excess observed by a detector in the Northern
Hemisphere that appears to continue into the sky regions accessible to IceCube,
although the excess identified as Region 1 in the IceCube
skymap~\citep{Abbasi:2011ai} is shifted to slightly lower right ascension
($122.4^\circ$).

A third excess region, Region C in Fig.\,\ref{fig:skymap_24h_multipolesub}, is
centered at $\alpha=205.7^\circ$ and $\delta=22.5^\circ$ with a pre-trial
significance of $8.2\,\sigma$ and a peak relative intensity of $(2.9 \pm 0.4
\pm 0.5)\times 10^{-4}$.  This excess region is not significant in the  Milagro
data, but the ARGO-YBJ collaboration has observed a hot spot at the same
location, called Region 4 in~\citet{ARGO-YBJ:2013gya}.  The morphology of this
region is shown in the right panel of Fig.\,\ref{fig:regionABC}.  The median
cosmic-ray energy at this declination is 2.0\,TeV.  

Region C is located at the center of the minimum of the large-scale structure,
but it is already visible (albeit with smaller significance) in
Fig.\,\ref{fig:skymap_24h_relint} before the subtraction of the
$\ell\leq3$ terms.  The relative intensity of this region in HAWC is a factor
of 1.8 higher than reported by ARGO-YBJ.

The significances quoted for the three excess regions do not account for
statistical trials caused by the search for any significant deviation from
isotropy in the 525\,716 pixels.  In a blind search we would account for
``look elsewhere'' effects by repeating the analysis for a large number of
isotropic sky maps with the same exposure as the data.  Because such a
calculation is computationally prohibitive given the high pre-trial
significance of the excess regions, we conservatively estimate that the number
of independent pixels in the sky map is of order $10^5$.  In fact, the trials
penalty is much smaller because we are not performing a blind search of the
data; these excess regions have been observed by other experiments.
However, even with a correction factor of $10^5$, the significances of Regions
A, B, and C are $16.1\sigma$, $10.2\sigma$, and $6.7\sigma$ after trials, respectively.

The ARGO-YBJ experiment has also observed a new region with a maximum relative
intensity of $2.3\times 10^{-4}$, called Region 3 in~\citet{ARGO-YBJ:2013gya},
which is a factor of 1.4 more intense than the excess in Region C.  The shape
of this new region is rather complex.  The most intense signal is found near
$\alpha=240^\circ$ and $\delta=45^\circ$, although the region extends to
declinations as low as $15^\circ$.  While this region is brighter in ARGO-YBJ
than Region C, it is currently not significant in HAWC data;  the largest
pre-trial significance within $10^\circ$ of the ARGO-YBJ peak excess is
$3.7\,\sigma$.  This region will be studied in more detail with a larger data
set in the future.

\subsection{Power Spectrum Analysis}
\label{subsec:powerspectrum}

A common tool to search for correlations between bins in a map without prior
knowledge of the expected angular scale of excess or deficit regions is the
angular power spectrum.  The amplitude of the power spectrum at multipole order
$\ell$ is correlated with the presence of structure at angular scales
$180^\circ/\ell$.  We perform an angular power spectrum analysis on the
unsmoothed relative intensity map $\delta I=\Delta N/\langle N\rangle$.  In
this analysis, $\delta I$ is treated as a scalar field which is expanded in
terms of a basis
\begin{linenomath}
\begin{equation}
\delta I(\alpha_i,\delta_i) =
    \sum_{\ell=0}^{\infty}
    \sum_{m=-\ell}^{\ell} a_{\ell m} Y_{\ell m}(\pi-\delta_i,\alpha_i)~~,
\end{equation}
\end{linenomath}
where the $Y_{\ell m}$ are the real (Laplace) spherical harmonics and the 
$a_{\ell m}$ are the multipole coefficients of the expansion in the sky map.  
The power spectrum of the relative intensity is defined as the variance of the 
multipole coefficients $a_{\ell m}$,
\begin{linenomath}
\begin{equation}
  {\cal C}_{\ell} = \frac{1}{2 \ell + 1} \sum_{m=-\ell}^{\ell} | a_{\ell m} |^{2}~~.
\end{equation}
\end{linenomath}

Due to the partial sky coverage of HAWC, the $Y_{lm}$ do not form an
orthonormal basis and the true power spectrum cannot be calculated directly.
Following the approach outlined in detail in~\citet{Abbasi:2011ai}, we first
calculate the so-called pseudo-power spectrum, a convolution of the power
spectrum of the data and the power spectrum of the corresponding relative
exposure map.  We use the publicly available PolSpice
software~\citep{Szapudi:2000xj,Chon:2003gx} to calculate the true power
spectrum from the pseudo-power spectrum.

\begin{figure*}[t]
  \centering
  \includegraphics[width=0.9\textwidth]{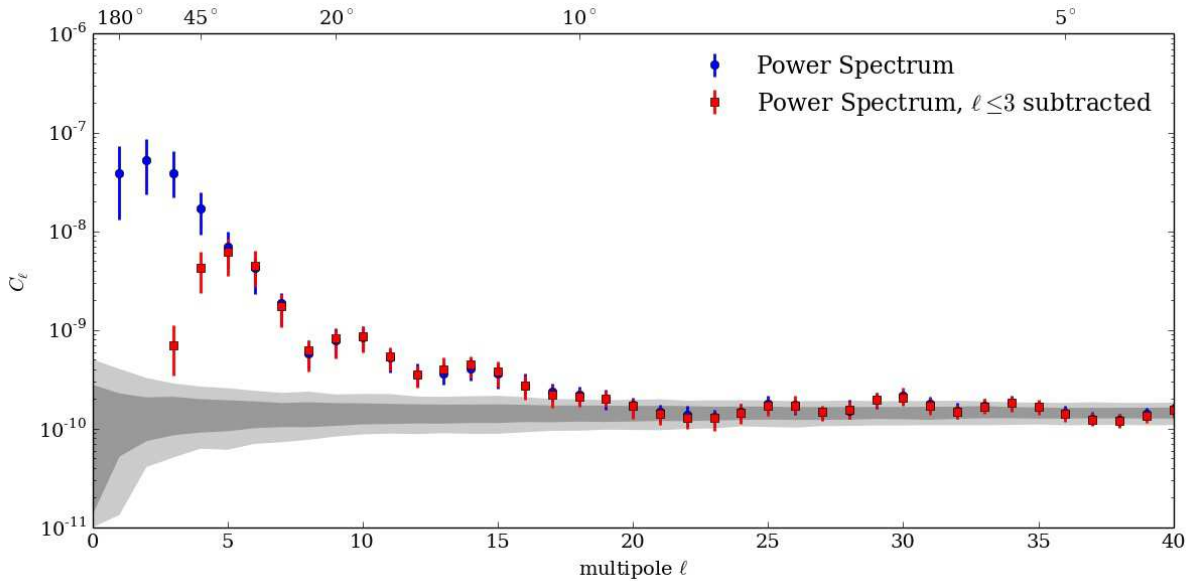}
  \caption{Angular power spectra of the unsmoothed relative intensity map
  (Fig.\,\ref{fig:skymap_24h_relint}) before (blue) and after (red) fitting 
  and subtraction of the dipole, quadrupole, and octupole moments ($\ell\leq3$).  
  The error bars on the $C_\ell$ are statistical.  Note that the $\ell<3$ 
  terms in the residual spectrum are not shown because they were found to be 
  compatible with zero within statistical uncertainties.  The gray bands show the 
  68\% and 95\% spread of the $C_\ell$ for isotropic data sets.}
  \label{fig:powerspectrum}
\end{figure*}

The angular power spectrum of the unsmoothed relative intensity map is shown
in Fig.\,\ref{fig:powerspectrum}.  The blue and red points show the power
spectrum before and after the subtraction of the $\ell\leq 3$ terms.  
The error bars are calculated from the diagonal components of the covariance 
matrix (see~\citet{Efstathiou:2003dj} for a detailed discussion).
The gray bands in Fig.\,\ref{fig:powerspectrum} 
indicate the 68\% and 95\% spread of the $C_\ell$ around the median for a large 
number of relative intensity maps representing isotropic arrival direction 
distributions.  These isotropic skymaps were generated by comparing the counts 
from the reference map to a Poisson-fluctuated reference map.  

The angular power spectrum of the relative intensity map shows, as expected, a
strong dipole $(\ell=1$) and quadrupole $(\ell=2)$ moment.  With increasing
$\ell$, the strength of the corresponding moments $C_{\ell}$ decreases, but
higher order multipoles up to $\ell=15$ still contribute significantly to the
sky map.  After subtraction of the dipole, quadrupole, and octupole ($\ell=3$) 
moments by the fit method described above, the dipole and quadrupole moments 
are missing in the spectrum and the octupole moment is diminished by two orders
of magnitude. All other moments are still present and, excluding $\ell=4$, have the 
same strength as in the original map given statistical uncertainties.
This indicates that the procedure described above is successful in reducing the
correlation between the different $\ell$ modes caused by the incomplete sky
coverage. However, the fact that the octupole moment is not completely removed 
after the fit shows that some correlation between modes persists.

As mentioned in Section\,\ref{subsec:map}, sky maps produced with the direct 
integration  method to estimate the reference level are potentially biased 
because the method can mask or reduce the strength of declination-dependent 
structures.  Since the angular power spectrum is based on these sky maps, it 
is also affected by this limitation of the technique.  The effect can lead to 
an underestimation of the power in certain multipoles, especially those with 
low $\ell$, and might thus distort the shape of the power spectrum.  It also 
complicates comparisons between the measured power spectrum and theoretical 
predictions.  However, the angular power spectrum remains a powerful diagnostic 
tool, for example in the evaluation of the two methods used to eliminate 
large-scale structure described in Section\,\ref{subsec:map}.

\subsection{Study of the Region A Excess}
\label{subsec:excess}

\begin{figure}[t]
  \centering
  \includegraphics[width=0.45\textwidth]{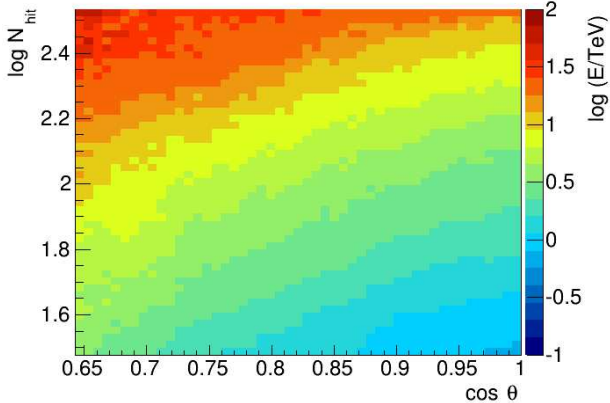}
  \caption{Median energy as a function of the number of triggered PMTs in the event,
   $N_{\mathrm{hit}}$, and the cosine of the zenith angle $\theta$ of the incident 
   cosmic ray, from simulation.}
  \label{fig:energyproxy}
\end{figure}

The study of Region A in Milagro data showed that the spectrum of the
cosmic-ray flux in this region is harder than the isotropic cosmic-ray flux,
with a possible cutoff around 10\,TeV.  At this point, a detailed study of the
energy dependence of the flux in the excess regions with HAWC is not possible.
Energy estimators based on the tank signal as a function of distance to the
shower core are currently being developed, but these techniques will only reach
their full potential with data from the complete 300-tank detector.  Here, we
perform a study based on a simple energy proxy that is based on the number of
PMTs  in the event and the zenith angle of the cosmic ray.  In 
Fig.\,\ref{fig:energyproxy}, we show the median cosmic-ray energy as a function of
these two parameters, based on simulations.  As expected, for a fixed number of
PMTs, the median energy rises with zenith angle, as the shower has to traverse
a larger integrated atmospheric depth.  

Based on this plot, we identify 7 bins in median energy 
given by
$(1.7^{+6.6}_{-1.3})$\,TeV, 
$(3.2^{+10.9}_{-2.4})$\,TeV, 
$(5.6^{+14.2}_{-3.9})$\,TeV, 
$(8.4^{+20.3}_{-5.9})$\,TeV, 
$(9.8^{+24.8}_{-6.7})$\,TeV, 
$(14.1^{+28.7}_{-9.9})$\,TeV, 
and $(19.2^{+32.3}_{-13.3})$\,TeV, 
respectively.  We define Region A as all pixels within a radius of $10^{\circ}$ 
about the center at $(\alpha,\delta)=(60.0^\circ,-7.1^\circ)$.  The relative 
intensity of the cosmic-ray flux in Region A is then obtained using the sum of 
all the angular bins in this region, for the 7 median energy bins.  To check the 
technique we also use the amplitude of a two-dimensional Gaussian fit to the 
relative intensity map.  Since the relative intensity of the excess as a function 
of radial distance to the center is relatively flat near the center, the methods 
give similar results.

\begin{figure}[t]
  \centering
  \includegraphics[width=0.45\textwidth]{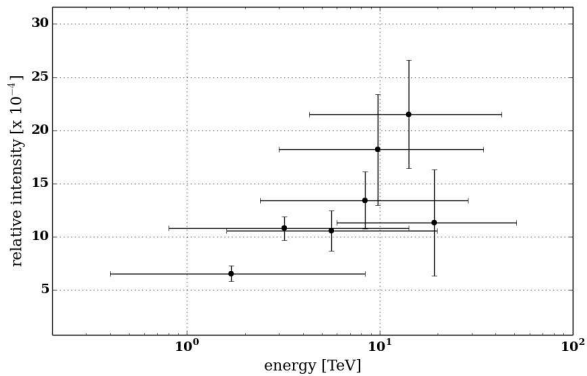}
  \caption{Spectrum of Region A in relative intensity in different energy proxy bins.
   The energies of the data were determined from Fig.~\ref{fig:energyproxy}.  The error bars
   on the median energy values correspond to a 68\% containing interval.}
  \label{fig:spectrum-regionA}
\end{figure}

The relative intensity of the flux in the excess regions is plotted as a
function of energy in Fig.\,\ref{fig:spectrum-regionA}.  The abscissae show the
median energy of each of the 7 bins, and the error bars correspond to the
68\% containing interval of each bin.  Despite the considerable overlap in
energy between the bins, the analysis is sufficient to confirm that the energy
spectrum of Region A is harder than the isotropic cosmic-ray spectrum.  

We estimate the statistical significance of the hard spectrum in Region A by
comparing the slope of a linear fit of $\delta I$ versus $\log{(E)}$ in
Fig.\,\ref{fig:spectrum-regionA} to similar fits performed at many random
locations in the field of view.  These locations excluded the $15^\circ$ circle
centered on Regions A, B, and C.  The distribution of slopes for the random locations
follows an approximately Gaussian distribution centered at zero with a width of
$1.2\times 10^{-4}$.  The slope at the position of Region A is $(4.5\pm 1.0)
\times 10^{-4}$, $3.8\,\sigma$ away from the mean.

\begin{figure*}[t]
  \centering
  \includegraphics[width=0.45\textwidth]{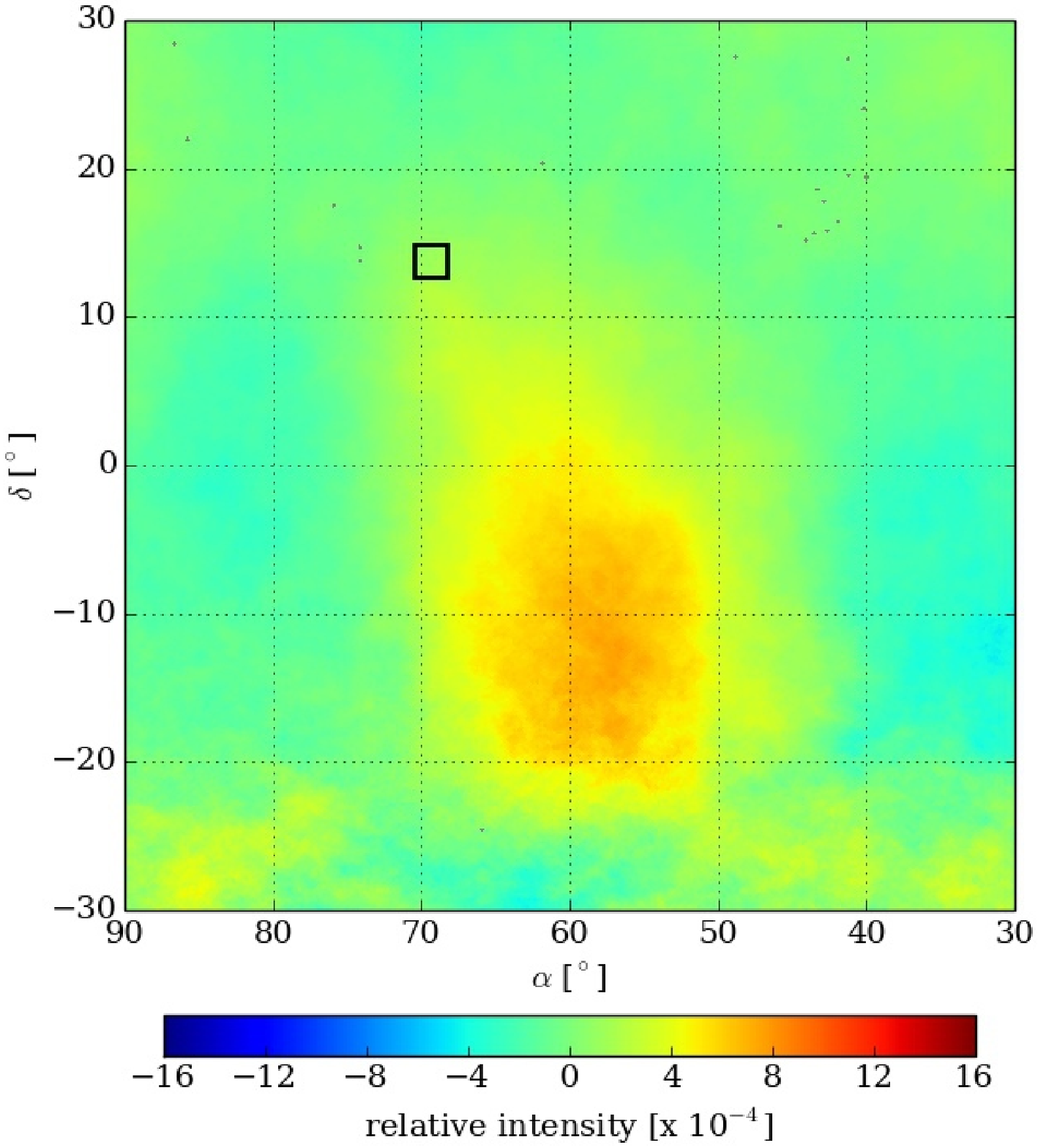}
  \includegraphics[width=0.45\textwidth]{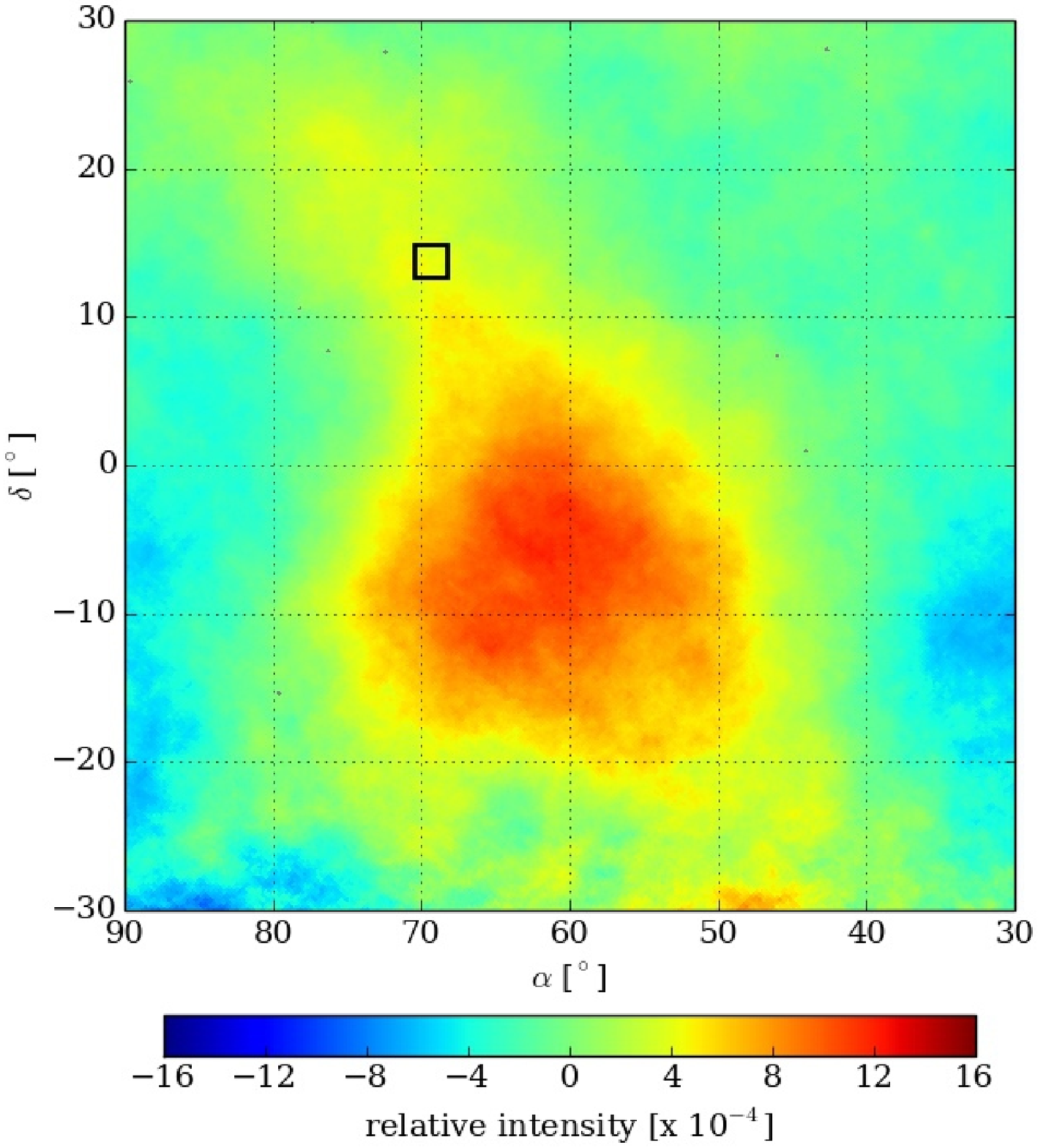}
  \includegraphics[width=0.45\textwidth]{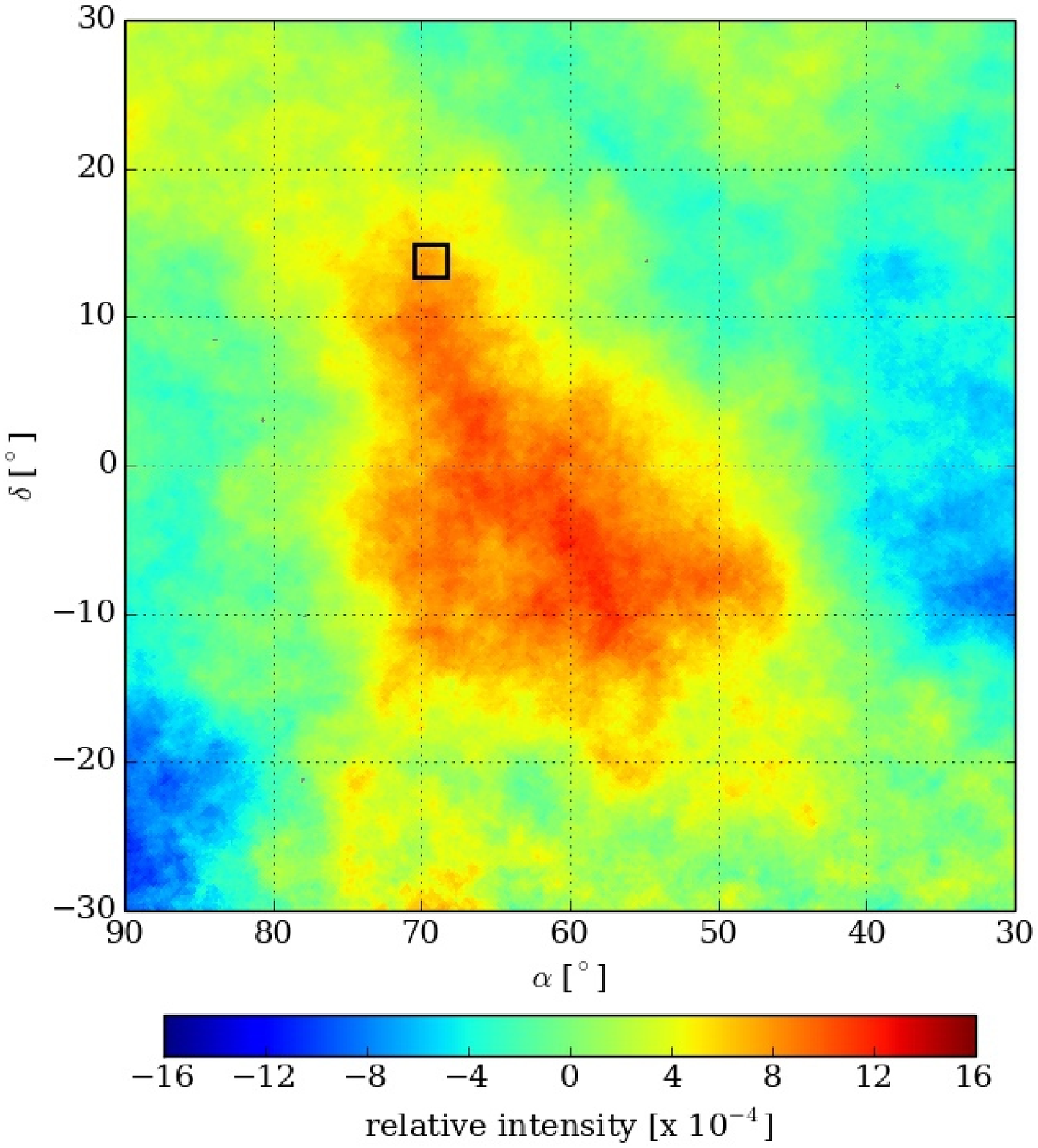}
  \includegraphics[width=0.45\textwidth]{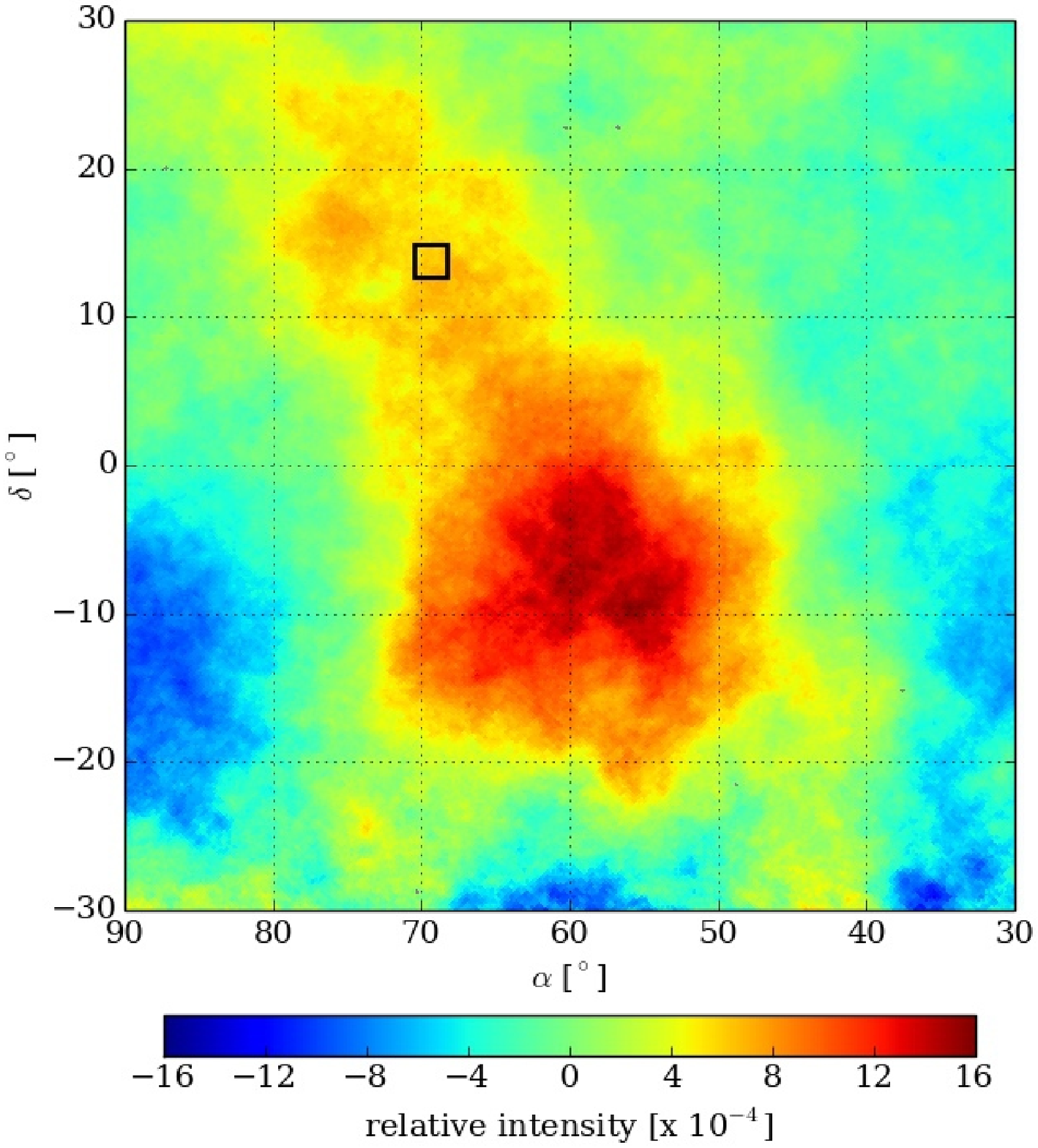}
  \caption{Relative intensity of Region A for 4 different energy proxy bins. The square mark 
  denotes the location of the centroid of Region A as reported by Milagro ($\alpha=69.4$, 
  $\delta=13.8$).  The median energy of the data in each plot is ${1.7}^{+6.6}_{-1.3}$\,TeV 
  {\slshape (top left)}, ${3.2}^{+10.9}_{-2.4}$\,TeV {\slshape (top right)},
  ${5.6}^{+14.2}_{-3.9}$\,TeV {\slshape (bottom left)}, and
  ${14.1}^{+28.7}_{-9.9}$\,TeV {\slshape (bottom right)}.
  }
  \label{fig:regionA-energy}
\end{figure*}

The relative intensity of Region A is plotted in several energy bins in
Fig.\,\ref{fig:regionA-energy}.  The four highest energy bins from
Fig.\,\ref{fig:spectrum-regionA} have been combined to boost the statistics of
the highest-energy plot.  The location of the centroid of Region A reported by
Milagro is plotted as a square marker.  The data indicate that Region A
changes in intensity and shape as a function of energy.  In the bin with the
lowest median energy, HAWC observes no significant excess at the location of
the Milagro centroid.  As the median energy increases, the relative intensity
of the excess observed in HAWC increases near the Milagro centroid.  In the two
bins of highest median energy, both measurements agree within uncertainties.

In their study of the energy dependence of the excess, the ARGO-YBJ
collaboration observed a similar effect~\citep{ARGO-YBJ:2013gya}.  The ARGO-YBJ
analysis splits Region A in two parts, an upper part roughly coinciding with
the brightest area in the Milagro map, and a lower part coinciding with the
HAWC excess.  At low energies, the lower part dominates, but as the energy
increases the upper region becomes as bright as the lower region.

The study of the morphology and relative intensity of all excess regions as a
function of energy will be continued with more data in the near future.  The
complete HAWC array will also have an improved energy resolution which will
allow for a cleaner binning of data as a function of energy.

\section{Conclusions}

Using $4.9\times 10^{10}$ events recorded with partial HAWC configurations
of 95 and 111 water-Cherenkov detectors we have observed a significant 
small-scale anisotropy in the arrival direction distribution of cosmic rays 
in the TeV band.  The observations are largely in agreement with previous 
measurements of the anisotropy in the Northern Hemisphere.  The sky map shows 
three regions of significantly enhanced cosmic-ray flux.  The two most 
significant excess regions (Regions A and B) coincide with regions that have also
been observed with the Milagro~\citep{Abdo:2008kr}, 
Tibet AS$\gamma$~\citep{Amenomori:2005dy}, 
and ARGO-YBJ experiments~\citep{ARGO-YBJ:2013gya}.  Discrepancies between 
experiments in the location and the relative intensity of the excess regions 
may be due to the presence of unaccounted energy effects in the anisotropy.  
We also confirm the presence of a third region of cosmic-ray excess (Region C)
which is not present in Milagro data, but was recently observed with 
ARGO-YBJ~\citep{ARGO-YBJ:2013gya}.  

Applying an energy estimator that is based on the number of PMTs in the event 
and the zenith angle of the cosmic ray, we also study the energy dependence of 
the relative intensity in the region of the most significant cosmic-ray excess 
(Region A).  We find that the spectrum in this region is harder than the 
isotropic cosmic-ray spectrum, in agreement with previous observations by Milagro
and ARGO-YBJ.

General features of the cosmic-ray arrival direction distribution in the
northern sky are also present in the southern sky, where the IceCube neutrino
observatory is currently the only experiment contributing to cosmic-ray
measurements in this energy band.  A combined analysis of data from both
hemispheres, with special attention to the small declination range where the
cosmic-ray sky is visible (at large zenith angles) with both IceCube and HAWC
will be performed in the near future.  Since HAWC observes almost all charged 
secondary particles in air showers and IceCube can observe only the muonic component 
that reaches the detector after about a mile of ice, a comparison of data in the 
overlap region might also give some insight into possible systematic effects. \\

We gratefully acknowledge Scott DeLay for his dedicated efforts in the 
construction and maintenance of the HAWC experiment.  This work has been 
supported by: 
the US National Science Foundation (NSF), 
the US Department of Energy Office of High-Energy Physics,
the Laboratory Directed Research and Development (LDRD) 
program of Los Alamos National Laboratory, 
Consejo Nacional de Ciencia y Tecnolog\'{\i}a (CONACyT), Mexico  
(grants 55155, 105666, 122331 and 132197), 
Red de F\'{\i}sica de Altas Energ\'{\i}as, Mexico, 
DGAPA-UNAM (grants IG100414-3, IN108713 and IN121309, IN115409, IN113612), 
VIEP-BUAP (grant 161-EXC-2011), 
the University of Wisconsin Alumni Research Foundation, 
the Institute of Geophysics, Planetary Physics, and 
Signatures at Los Alamos National Laboratory, and
the Luc Binette Foundation UNAM Postdoctoral Fellowship program.

\bibliography{hcr} 

\end{document}

%% file: authorsapj.tex
\author{A.~U.~Abeysekara\altaffilmark{1,6},
R.~Alfaro\altaffilmark{2},
C.~Alvarez\altaffilmark{3},
J.~D.~{\'A}lvarez\altaffilmark{4},
R.~Arceo\altaffilmark{3},
J.~C.~Arteaga-Vel{\'a}zquez\altaffilmark{4},
H.~A.~Ayala~Solares\altaffilmark{5},
A.~S.~Barber\altaffilmark{6},
B.~M.~Baughman\altaffilmark{7},
N.~Bautista-Elivar\altaffilmark{8},
E.~Belmont\altaffilmark{2},
S.~Y.~BenZvi\altaffilmark{10,17},
D.~Berley\altaffilmark{7},
M.~Bonilla~Rosales\altaffilmark{11},
J.~Braun\altaffilmark{7,17},
K.~S.~Caballero-Mora\altaffilmark{13},
A.~Carrami{\~n}ana\altaffilmark{11},
M.~Castillo\altaffilmark{14},
U.~Cotti\altaffilmark{4},
J.~Cotzomi\altaffilmark{14},
E.~de~la~Fuente\altaffilmark{15},
C.~De~Le{\'o}n\altaffilmark{4},
T.~DeYoung\altaffilmark{1},
R.~Diaz~Hernandez\altaffilmark{11},
J.~C.~D{\'\i}az-V{\'e}lez\altaffilmark{17},
B.~L.~Dingus\altaffilmark{18},
M.~A.~DuVernois\altaffilmark{17},
R.~W.~Ellsworth\altaffilmark{19,7},
D.~W.~Fiorino\altaffilmark{17,*},
N.~Fraija\altaffilmark{20},
A.~Galindo\altaffilmark{11},
F.~Garfias\altaffilmark{20},
M.~M.~Gonz{\'a}lez\altaffilmark{20},
J.~A.~Goodman\altaffilmark{7},
M.~Gussert\altaffilmark{21},
Z.~Hampel-Arias\altaffilmark{17},
J.~P.~Harding\altaffilmark{18},
P.~H{\"u}ntemeyer\altaffilmark{5},
C.~M.~Hui\altaffilmark{5},
A.~Imran\altaffilmark{18,17},
A.~Iriarte\altaffilmark{20},
P.~Karn\altaffilmark{30,17},
D.~Kieda\altaffilmark{6},
G.~J.~Kunde\altaffilmark{18},
A.~Lara\altaffilmark{12},
R.~J.~Lauer\altaffilmark{22},
W.~H.~Lee\altaffilmark{20},
D.~Lennarz\altaffilmark{23},
H.~Le{\'o}n~Vargas\altaffilmark{2},
J.~T.~Linnemann\altaffilmark{1},
M.~Longo\altaffilmark{21},
R.~Luna-Garc{\'\i}a\altaffilmark{24},
K.~Malone\altaffilmark{16},
A.~Marinelli\altaffilmark{2},
S.~S.~Marinelli\altaffilmark{1},
H.~Martinez\altaffilmark{13},
O.~Martinez\altaffilmark{14},
J.~Mart{\'\i}nez-Castro\altaffilmark{24},
J.~A.~J.~Matthews\altaffilmark{22},
J.~McEnery\altaffilmark{9},
E.~Mendoza~Torres\altaffilmark{11},
P.~Miranda-Romagnoli\altaffilmark{25},
E.~Moreno\altaffilmark{14},
M.~Mostaf{\'a}\altaffilmark{16},
L.~Nellen\altaffilmark{26},
M.~Newbold\altaffilmark{6},
R.~Noriega-Papaqui\altaffilmark{25},
T.~Oceguera-Becerra\altaffilmark{15,2},
B.~Patricelli\altaffilmark{20},
R.~Pelayo\altaffilmark{24,31},
E.~G.~P{\'e}rez-P{\'e}rez\altaffilmark{8},
J.~Pretz\altaffilmark{16},
C.~Rivi{\`e}re\altaffilmark{20,7},
D.~Rosa-Gonz{\'a}lez\altaffilmark{11},
E.~Ruiz-Velasco\altaffilmark{2},
J.~Ryan\altaffilmark{27},
H.~Salazar\altaffilmark{14},
F.~Salesa~Greus\altaffilmark{16},
A.~Sandoval\altaffilmark{2},
M.~Schneider\altaffilmark{28},
G.~Sinnis\altaffilmark{18},
A.~J.~Smith\altaffilmark{7},
K.~Sparks~Woodle\altaffilmark{16},
R.~W.~Springer\altaffilmark{6},
I.~Taboada\altaffilmark{23},
P.~A.~Toale\altaffilmark{29},
K.~Tollefson\altaffilmark{1},
I.~Torres\altaffilmark{11},
T.~N.~Ukwatta\altaffilmark{1,18},
L.~Villase{\~n}or\altaffilmark{4},
T.~Weisgarber\altaffilmark{17},
S.~Westerhoff\altaffilmark{17},
I.~G.~Wisher\altaffilmark{17},
J.~Wood\altaffilmark{7},
G.~B.~Yodh\altaffilmark{30},
P.~W.~Younk\altaffilmark{18},
D.~Zaborov\altaffilmark{16},
A.~Zepeda\altaffilmark{13},
and H.~Zhou\altaffilmark{5}
\newline (The HAWC Collaboration)}

\altaffiltext{1}{Department of Physics \& Astronomy, Michigan State University, East Lansing, MI, USA}
\altaffiltext{2}{Instituto de F{\'\i}sica, Universidad Nacional Aut{\'o}noma de M{\'e}xico, Mexico D.F., Mexico}
\altaffiltext{3}{CEFyMAP, Universidad Aut{\'o}noma de Chiapas, Tuxtla Guti{\'e}rrez, Chiapas, Mexico}
\altaffiltext{4}{Universidad Michoacana de San Nicol{\'a}s de Hidalgo, Morelia, Mexico}
\altaffiltext{5}{Department of Physics, Michigan Technological University, Houghton, MI, USA}
\altaffiltext{6}{Department of Physics \& Astronomy, University of Utah, Salt Lake City, UT, USA}
\altaffiltext{7}{Department of Physics, University of Maryland, College Park, MD, USA}
\altaffiltext{8}{Universidad Polit{\'e}cnica de Pachuca, Pachuca, Hidalgo, Mexico}
\altaffiltext{9}{NASA Goddard Space Flight Center, Greenbelt, MD, USA}
\altaffiltext{10}{Department of Physics \& Astronomy, University of Rochester, Rochester, NY, USA}
\altaffiltext{11}{Instituto Nacional de Astrof{\'\i}sica, {\'O}ptica y Electr{\'o}nica, Tonantzintla, Puebla, Mexico}
\altaffiltext{12}{Instituto de Geof{\'\i}sica, Universidad Nacional Aut{\'o}noma de M{\'e}xico, Mexico D.F., Mexico}
\altaffiltext{13}{Centro de Investigaci{\'o}n y de Estudios Avanzados del Instituto Polit{\'e}cnico Nacional, Mexico D.F., Mexico}
\altaffiltext{14}{Facultad de Ciencias F{\'\i}sico Matem{\'a}ticas, Benem{\'e}rita Universidad Aut{\'o}noma de Puebla, Puebla, Mexico}
\altaffiltext{15}{IAM-Dpto. de Fisica; Dpto. de Electronica (CUCEI), IT.Phd (CUCEA), Phys\_Mat. Phd (CUVALLES), Universidad de Guadalajara, Jalisco, Mexico}
\altaffiltext{16}{Department of Physics, Pennsylvania State University, University Park, PA, USA}
\altaffiltext{17}{Wisconsin IceCube Particle Astrophysics Center (WIPAC) \& Department of Physics, University of Wisconsin-Madison, Madison, WI, USA}
\altaffiltext{18}{Physics Division, Los Alamos National Laboratory, Los Alamos, NM, USA}
\altaffiltext{19}{School of Physics, Astronomy \& Computational Sciences, George Mason University, Fairfax, VA, USA}
\altaffiltext{20}{Instituto de Astronom{\'\i}a, Universidad Nacional Aut{\'o}noma de M{\'e}xico, Mexico D.F., Mexico}
\altaffiltext{21}{Physics Department, Colorado State University, Fort Collins, CO, USA}
\altaffiltext{22}{Department of Physics \& Astronomy, University of New Mexico, Albuquerque, NM, USA}
\altaffiltext{23}{School of Physics \& Center for Relativistic Astrophysics, Georgia Institute of Technology, Atlanta, GA, USA}
\altaffiltext{24}{Centro de Investigaci{\'o}n en Computaci{\'o}n, Instituto Polit{\'e}cnico Nacional, Mexico D.F., Mexico}
\altaffiltext{25}{Universidad Aut{\'o}noma del Estado de Hidalgo, Pachuca, Hidalgo, Mexico}
\altaffiltext{26}{Instituto de Ciencias Nucleares, Universidad Nacional Aut{\'o}noma de M{\'e}xico, Mexico D.F., Mexico}
\altaffiltext{27}{Space Science Center, University of New Hampshire, Durham, NH, USA}
\altaffiltext{28}{Santa Cruz Institute for Particle Physics, University of California, Santa Cruz, Santa Cruz, CA, USA}
\altaffiltext{29}{Department of Physics \& Astronomy, University of Alabama, Tuscaloosa, AL, USA}
\altaffiltext{30}{Department of Physics \& Astronomy, University of California, Irvine, Irvine, CA, USA}
\altaffiltext{31}{Unidad Profesional Interdisciplinaria de Ingenier{\'i}a y Tecnolog{\'i}as Avanzadas del Instituto Polit{\'e}cnico Nacional, M{\'e}xico, D.F., Mexico}
\altaffiltext{*}{dan.fiorino@wipac.wisc.edu}